\documentclass{epsv8}

\usepackage{amsmath}
\usepackage{bm}

\usepackage{caption}

\usepackage{graphicx}

\title{Numerical Investigation on the Compressive Behavior of Hierarchical Granular Piles}
\author{
Sota Arakawa$^{*}$,
Japan Agency for Marine-Earth Science and Technology, 3173-25, Showa-machi, Kanazawa-ku, Yokohama 236-0001, Japan,
arakawas@jamstec.go.jp\\
Mikito Furuichi,
Japan Agency for Marine-Earth Science and Technology, m-furuic@jamstec.go.jp \\
Daisuke Nishiura,
Japan Agency for Marine-Earth Science and Technology, nishiura@jamstec.go.jp 
}

\abstract{
Hierarchical granular piles composed of aggregates are key structural features in both geoscience and planetary science, from fault gouge in seismic zones to the internal structures of comets.
Although experimental studies have suggested a multi-step evolution in their packing structure, this hypothesis has lacked numerical validation.
In this study, we performed large-scale numerical simulations using the discrete element method to investigate the compressive behavior of hierarchical granular piles.
We successfully reproduced and confirmed a three-stage evolution process: (i) rearrangement of the aggregate packing structure, (ii) plastic deformation of small aggregates, and (iii) elastic deformation of constituent particles.
Additionally, we developed a semi-analytical model for the compression curve, offering insights into the compressive stages and structural dynamics.
Our findings have applications in modeling the internal density profiles of comets and in understanding the early thermal evolution of small icy bodies.
}


\begin{document}

\maketitle

\section{Introduction}

Granular matter is ubiquitous on Earth, and the inelastic deformation of macroscopic grains within granular piles plays a key role in many geophysical phenomena.
For example, fault gouge, a layer of fragmented rock, is often observed along fault sections \citep{2005Natur.434..749W}.
The physical state of fault gouge controls the frictional strength and stability of faults, making its development a critical process in earthquake dynamics \citep{1975PApGe.113...69E, 2005GeoRL..32.5305A, 2009GeoRL..3623302A, 2022Natur.606..922R}.
Similarly, sediment layers of crushable volcanic pumice contribute to rainfall- and earthquake-induced slope failures \citep{2024SoFou..6401465S}, and understanding their mechanical properties is essential for evaluating the potential for these disasters.
Furthermore, natural soil also has a hierarchical structure, where fine clay grains form small aggregates using sticky organic substances \citep{YASUDA2023103896, YUDINA2023116478}.
The compression of hierarchical soil alters its water retention and air exchange capacity, thus the compressive behavior of hierarchical granular piles is a key factor for agriculture, planetary habitability, and other fields.

Hierarchical granular piles have garnered attention not only in the community of geoscience but also in planetary science.
For instance, comets are thought to be loose granular piles composed of porous dust aggregates \citep{2012Icar..221....1S, 2020MNRAS.493.3690G, 2023MNRAS.523.5841C}.
In the solar protoplanetary disk, micron-sized interstellar dust particles could grow into macroscopic aggregates via pairwise collisions when the aggregates are small \citep{2008ARA&A..46...21B, 2021ApJ...915...22H}.
In contrast, both numerical simulations \citep{2023ApJ...951L..16A, 2025ApJ...983...75O} and laboratory experiments \citep{2013Icar..225...75K} have revealed that the sticking probability upon collision decreases as the aggregate radius increases.
Additionally, continued non-sticking collisions lead to rounding and compaction of aggregates \citep{2009ApJ...696.2036W}.
These aggregates can accumulate locally via aerodynamical processes, and comets would be formed via the gravitational collapse of dust clumps \citep{2017MNRAS.469S.755B}.
This scenario suggests that the interior structure of comets is a hierarchical granular pile, which is consistent with the weak mechanical strengths \citep{2014Icar..235..156B, 2019A&A...630A...2H} and low thermal inertia \citep{2020MNRAS.497.1166A} reported by ground-based observations and by the {\it Rosetta} space mission.
Since the internal density structure of small celestial bodies is determined by the balance between self-gravity and compressive strength \citep{2018ApJ...860..123O, 2024ApJ...974....9T}, studying the compressive behavior of hierarchical granular piles could provide important insights into the origin of comets in the solar system.

\begin{figure}
\centering
\includegraphics[width = 0.6\columnwidth]{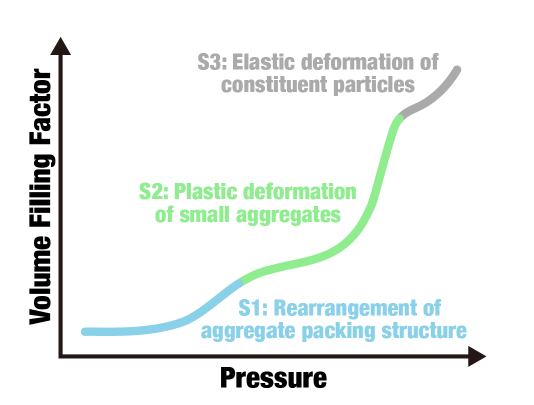}
\caption{
Schematic of the compression curve of hierarchical granular piles.
The compression curve can be divided into three stages: rearrangement of the aggregate packing structure (Stage 1), plastic deformation of small aggregates (Stage 2), and elastic deformation of constituent particles (Stage 3).
}
\label{fig:schematic_curve}
\end{figure}

The compressive behavior of hierarchical granular piles has been investigated primarily through experimental studies \citep{2015Icar..257...33S, 2021PhRvR...3a3190P, 2022MNRAS.514.3366M}.
Several studies have also conducted numerical simulations of the compression of hierarchical granular piles consisting of crushable aggregates \citep{ShingoIshihara201855.492, 2023NatSR..13.7843C}.
These studies have reported that the packing structure of hierarchical granular piles evolves during compression in a multi-step process (see Figure \ref{fig:schematic_curve}).
Under low confining pressure, the deformation is characterized by the rotation of aggregates, whereas under high confining pressure, aggregate breakage drives structural evolution.
Furthermore, deformation of constituent particles occurs when the applied pressure is sufficiently high.

However, the compression behavior of hierarchical granular piles and its dependence on material properties are still not fully understood in detail.
For instance, the rolling motion of aggregates within hierarchical granular piles is controlled by inter-aggregate friction, which would be influenced by the inter-aggregate cohesion.
Nevertheless, to the best of our knowledge, there are no reliable quantitative models that can predict the relationship between the rolling motion of aggregates and their material and structural properties.

In this study, we perform numerical simulations of the compression behavior of hierarchical granular piles composed of aggregates (see Figure \ref{fig:snapshot}).
We analyze the compressive behavior of these piles from a multi-scale perspective.
Our results show that the packing structure evolves through a three-step process.
Additionally, we develop a quantitative semi-analytical model for the compression curve.
Our findings are broadly applicable to the deformation of hierarchical granular piles under confining pressure.

\section{Methods}
\label{sec:methods}

We perform numerical compression tests of hierarchical granular piles using the discrete element method (DEM) \citep{doi:10.1680/geot.1979.29.1.47}.
We briefly introduce the particle interaction models used in our DEM simulations (Section \ref{sec:DEM}) and describe the simulation setup (Section \ref{sec:setup}).
Given that one of the target applications of our granular study is the formation of comets, the simulation setup is designed for small celestial bodies primarily composed of ice.
We consider granular piles consisting of small aggregates with a radius of $r_{\rm agg}$, and each aggregate is made of spherical ice particle with a radius of $r_{\bullet} = 0.1~{\mu}{\rm m}$.
The material properties of ice particles are set to match those used in our previous studies \citep{2023A&A...670L..21A}; Young's modulus is $E = 7~{\rm GPa}$, Poisson's ratio is $\nu = 0.25$, the surface energy is $\gamma = 0.1~{\rm J}~{\rm m}^{-2}$, and the material density is $\rho_{\bullet} = 1000~{\rm kg}~{\rm m}^{-3}$.

\subsection{Particle interaction model}
\label{sec:DEM}

The DEM simulation solves the motion of each particle at every time step.
The normal force acting between two particles, $F$, is given by the sum of the two terms:
\begin{equation}
F = F_{\rm E} + F_{\rm D},
\end{equation}
where $F_{\rm E}$ denotes the force arising from the elastic deformation and $F_{\rm D}$ is the force related to the viscous dissipation.
The sign of $F$ is positive when the repulsive force acts on particles in contact.
We also calculate particle--wall interactions in the same manner.

We consider a contact of elastic spheres having a surface energy \citep{1971RSPSA.324..301J}.
Two particles in contact make a contact area.
The radius of the contact area, $a$, depends on the compression length, $\delta$.
For two particles in contact, $\delta$ is defined as $\delta = 2 r_{\bullet} - d$, where $d$ is the distance between the centers of the two particles.
For a particle--wall contact, $\delta$ is given by $\delta = r_{\bullet} - d_{\rm wall}$, where $d_{\rm wall}$ is the distance between the center of the particle and the surface of the wall.

We calculate $a$ and $F_{\rm E}$ using a particle interaction model derived by \cite{1971RSPSA.324..301J}.
At the equilibrium state where $F_{\rm E} = 0$, $\delta$ is equal to the equilibrium compression length, $\delta_{0}$, and $a$ is equal to the equilibrium radius, $a_{0}$.
The maximum force needed to separate the particles in contact is $F_{\rm crit}$.
The values of $\delta_{0}$, $a_{0}$, and $F_{\rm crit}$ are summarized in Table \ref{table:equilibrium}.

\begin{table}
\caption{The equilibrium compression length, $\delta_{0}$, the equilibrium radius, $a_{0}$, and the maximum force needed to separate the particles in contact, $F_{\rm crit}$, for spherical ice particles with a radius of $r_{\bullet} = 0.1~{\mu}{\rm m}$.}
\centering
\begin{tabular}{c c c}
\hline
    &  particle--particle  &  particle--wall   \\
\hline
   $\delta_{0}$ (nm)        &  $1.020$  &  $1.286$  \\  
   $a_{0}$ (nm)             &  $12.37$  &  $19.64$  \\  
   $F_{\rm crit}$ (nN)      &  $47.12$  &  $94.25$  \\  
\hline
\end{tabular}
\label{table:equilibrium}
\end{table}

The relationship between $a / a_{0}$ and $\delta / \delta_{0}$ is shown in Figure \ref{fig:JKR}(a).
Their relationship is given by solving the following equation:
\begin{equation}
\frac{\delta}{\delta_{0}} = 3 {\left( \frac{a}{a_{0}} \right)}^{2} - 2 {\left( \frac{a}{a_{0}} \right)}^{1/2}.
\label{eq:JKR_a}
\end{equation}
\cite{2024PhRvE.109b4904A} have derived an equivalent equation which explicitly expresses $a$ as a function of $\delta$ (see also \cite{CHEN2023118742}), and we use this explicit equation in our simulation code. 
Similarly, the relationship between $F_{\rm E} / F_{\rm crit}$ and $a / a_{0}$ is given by
\begin{equation}
\frac{F_{\rm E}}{F_{\rm crit}} = 4 {\left( \frac{a}{a_{0}} \right)}^{3} - 4 {\left( \frac{a}{a_{0}} \right)}^{3/2}.
\label{eq:JKR_F}
\end{equation}
The relationship between $F_{\rm E} / F_{\rm crit}$ and $\delta / \delta_{0}$ is shown in Figure \ref{fig:JKR}(b), which is obtained by combining Equations (\ref{eq:JKR_a}) and (\ref{eq:JKR_F}).
A contact forms at $\delta / \delta_{0} = 0$, while the contact breaks when $\delta / \delta_{0}$ reaches $- {( 9 / 16 )}^{1/3}$.

We also consider a viscous dissipation force to damp the oscillatory motion of particles.
We use a viscous dissipation model \citep{2013JPhD...46Q5303K, 2021ApJ...910..130A} and $F_{\rm D}$ is given as follows:
\begin{equation}
F_{\rm D} = C_{\rm D} a \dot{\delta},
\end{equation}
where $C_{\rm D} = 0.7168~{\rm kg}~{\rm m}^{-1}~{\rm s}^{-1}$ is the dissipation coefficient.
The value of $C_{\rm D}$ assumed here is equal to that used in \cite{2023A&A...670L..21A}.

In this study, we do not include tangential interactions (i.e., frictional torques by rolling, sliding, and twisting between two particles in contact).
Therefore, we only solve the translational motion of constituent particles and do not consider their rotational motion.

These interaction models are implemented in a DEM code that is optimized for large scale simulations using hybrid MPI and OpenMP parallelization, {\bf DEPTH} (DEM-based parallel multi-physics simulator) \citep{FURUICHI2017135, NISHIURA20214432, 2023Tectp.86229963F, 2024Tectp.87430230F}.
We utilized up to 26 million particles (see Section \ref{sec:setup}) using up to 64 CPUs (4096 cores) of the Earth Simulator (ES4) at Japan Agency for Marine-Earth Science and Technology (JAMSTEC).

\subsection{Simulation procedure}
\label{sec:setup}

We prepare a hierarchical granular pile consisting of small spherical aggregates as the initial condition (see Figure \ref{fig:initial_schematic}).
The spherical aggregates used in this study are generated using the close-packing and particle-extraction (CPE) procedure \citep{2011ApJ...737...36W, 2023ApJ...951L..16A}.
The packing fraction of the aggregates is set to $\phi_{\rm agg} = 0.5$, and the number of particls within each aggregate is $N_{\rm par} = \phi_{\rm agg} {( r_{\rm agg} / r_{\bullet} )}^{3}$.
We treat $r_{\rm agg}$ as a parameter and present numerical results for $r_{\rm agg} = 32 r_{\bullet}$ and $64 r_{\bullet}$ in this study.
The initial positions of the aggregates are determined by the random ballistic deposition (RBD) procedure \citep{2004PhRvL..93k5503B, 2011Icar..214..286K}, and the number of aggregates within the computational domain is set to $N_{\rm agg} = 200$.
Thus, the total number of particles is given by $N_{\rm tot} = N_{\rm agg} N_{\rm par}$, with $N_{\rm tot} = 3,276,800$ for $r_{\rm agg} = 32 r_{\bullet}$ and $N_{\rm tot} = 26,214,400$ for $r_{\rm agg} = 64 r_{\bullet}$, respectively.

Hierarchical granular piles are packed within a rectangular simulation box.
We use a periodic boundary condition for the four horizontal boundaries perpendicular to the direction of compression.
The top wall moves with a speed $v_{\rm wall}$ to compress the piles whereas the bottom wall fixes at the height $z = 0$.
The initial condition for the case of $r_{\rm agg} = 32 r_{\bullet}$ is illustrated in Figure \ref{fig:snapshot}(a).
The speed of the top wall is given by 
\begin{equation}
v_{\rm wall} = \frac{L_{\rm z}}{\tau_{\rm comp}},
\end{equation}
where $L_{\rm z}$ is the height of the box (i.e., the distance between the top and bottom walls) and $\tau_{\rm comp}$ is the compression timescale.
We adopted a constant $\tau_{\rm comp}$ following \citet{2013A&A...554A...4K}.
As $v_{\rm wall} = - {\rm d}L_{\rm z} / {\rm d}t$ by definition, the temporal change of $L_{\rm z}$ is given by $L_{\rm z} (t) = L_{\rm z, ini} \exp{\left( - t / \tau_{\rm comp} \right)}$, where $L_{\rm z, ini}$ is the height at $t = 0$.
We set $\tau_{\rm comp} = 2 \times 10^{-4}~{\rm s}$ for $r_{\rm agg} = 32 r_{\bullet}$ and $\tau_{\rm comp} = 4 \times 10^{-4}~{\rm s}$ for $r_{\rm agg} = 64 r_{\bullet}$.
Both the width and length of the box, $L_{\rm x}$ and $L_{\rm y}$, are the same and set to $L_{\rm x} = L_{\rm y} = 12 r_{\rm agg}$.
The volume of the box is given by $V_{\rm box} = L_{\rm x} L_{\rm y} L_{\rm z}$.
The volume filling factor of the hierarchical granular pile is defined as $\phi = N_{\rm tot} V_{\bullet} / V_{\rm box}$, where $V_{\bullet} = {( 4 \pi / 3 )} {r_{\bullet}}^{3}$ is the volume of each particle.
Thus, the relationship between $v_{\rm wall}$ and $\phi$ is given by $v_{\rm wall} = 0.31 \times {( \phi / 0.15 )}^{-1}~{\rm m}~{\rm s}^{-1}$.

\section{Compression test of hierarchical granular pile}
\label{sec:results}

In Section \ref{sec:results}, we present the simulation results of compression of hierarchical granular piles.
Additionally, we perform compression tests of single aggregate in Section \ref{sec:single}.
The semi-analytical model is derived in Section \ref{sec:semi-analytic}.
In this study, we introduce several types of ``filling factors," and quantities associated with $\phi$ are summarized in Table \ref{table:phi}.

\begin{table}
\caption{List of quantities associated with $\phi$.}
\centering
\begin{tabular}{l l}
\hline
Character  &  Definition   \\
\hline
   $\phi$               &  Volume filling factor of the hierarchical granular pile (Section \ref{sec:setup})  \\  
   $\phi_{\rm agg}$     &  Packing fraction of the aggregates ($\phi_{\rm agg} = 0.5$; Section \ref{sec:setup})  \\ 
   $\phi_{\rm str}$     &  Filling factor of the aggregate packing structure ($\phi_{\rm str} \equiv \phi / \phi_{\rm agg}$; Section \ref{sec:inter-aggregate})  \\  
   $\phi_{\rm sect} {( z )}$    &  Filling factor at the height $z$ (Section \ref{sec:structure_evolution})  \\  
\hline
\end{tabular}
\label{table:phi}
\end{table}

\subsection{Structure evolution during compression}
\label{sec:structure_evolution}

\begin{figure}
\centering
\includegraphics[width = \columnwidth]{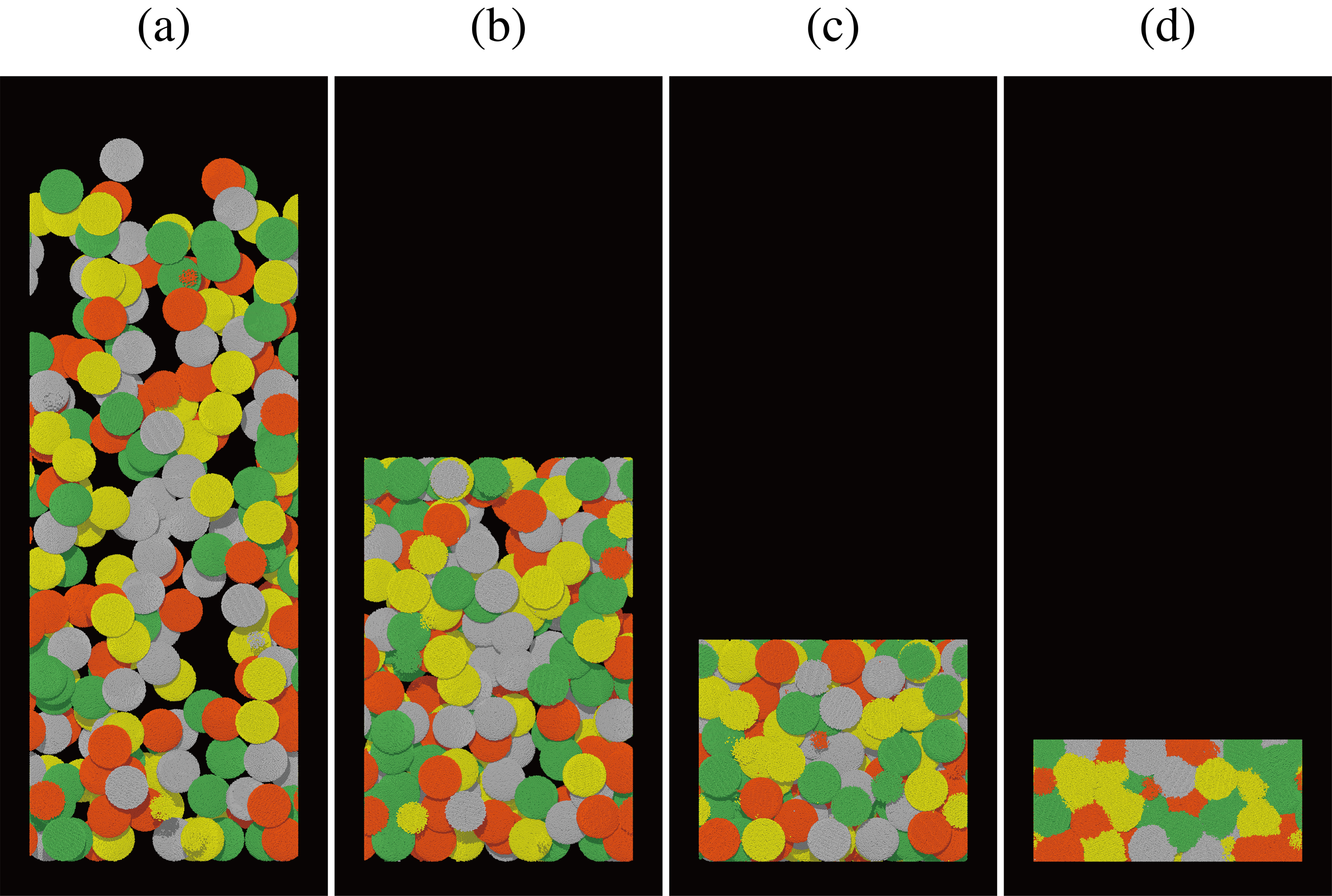}
\caption{
Snapshots of a hierarchical granular pile with $r_{\rm agg} = 32 r_{\bullet}$ during compression.
(a) Initial condition.
(b) $\phi = 0.160$. 
(c) $\phi = 0.292$. 
(d) $\phi = 0.532$. 
}
\label{fig:snapshot}
\end{figure}

\begin{figure}
\centering
\includegraphics[width = \columnwidth]{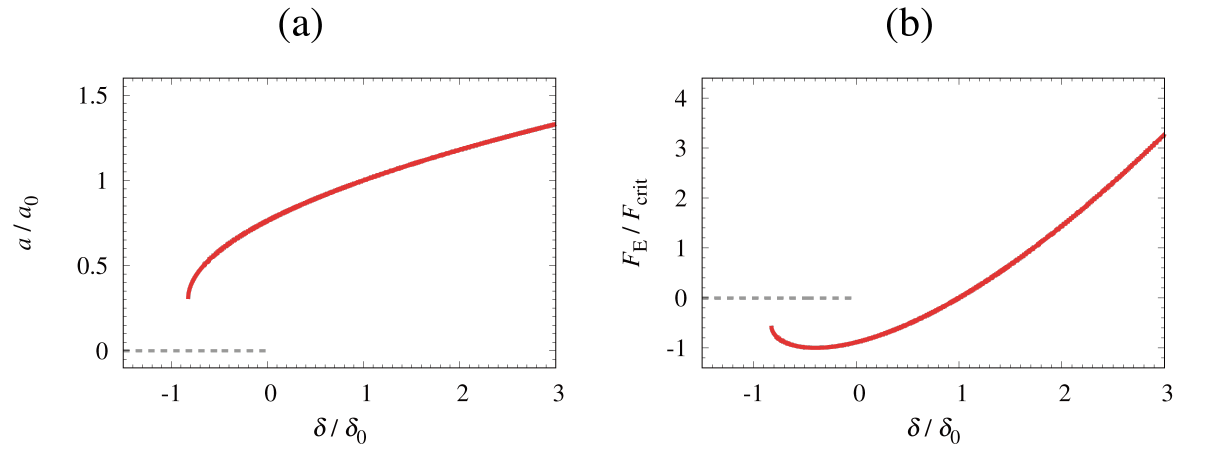}
\caption{
Contact radius, $a$, and the elastic term of the contact force, $F_{\rm E}$, with respect to the compression length, $\delta$.
(a) $a$ with respect to $\delta$.
(b) $F_{\rm E}$ with respect to $\delta$.
Red solid lines represent the relations for particles in contact, while gray dashed lines are for separated particles.
A contact initiates at $\delta / \delta_{0} = 0$ and it breaks at $\delta / \delta_{0} = - {( 9 / 16 )}^{1/3}$.
}
\label{fig:JKR}
\end{figure}

\begin{figure}
\centering
\includegraphics[width = 0.4\columnwidth]{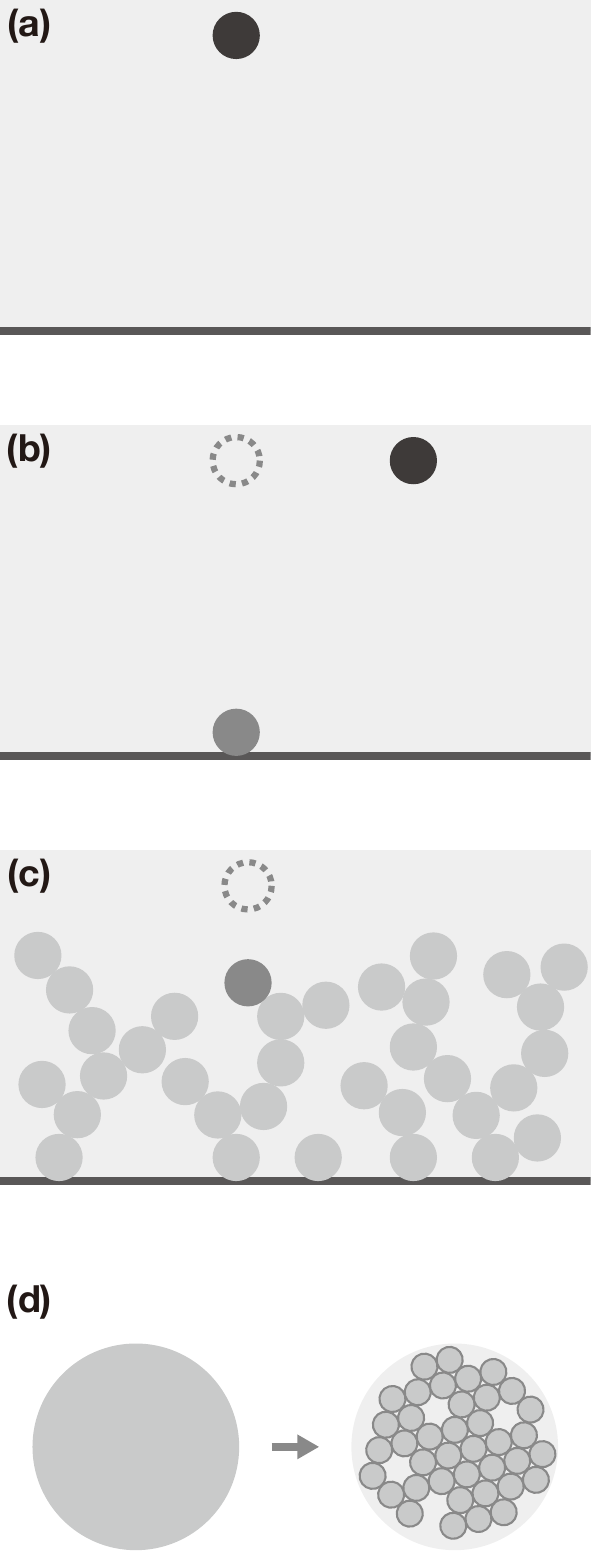}
\caption{
Schematic of the preparation procedure of hierarchical granular piles.
(a) We deposit a sphere with a radius $r_{\rm agg}$ on the bottom wall from a random position.
(b) We sequentially repeat the deposition process.
The motion of the sphere is terminated when it collide with another sphere or the bottom wall.
(c) We prepare an aggregate consisting of 200 spheres by the random ballistic deposition (RBD) procedure.
(d) We replace spheres with aggregates made by the close-packing and particle-extraction (CPE) procedure.
}
\label{fig:initial_schematic}
\end{figure}

Figure \ref{fig:snapshot} shows snapshots of a hierarchical granular pile with $r_{\rm agg} = 32 r_{\bullet}$ during compression.
Each panel corresponds to (a) the initial state, (b) $\phi = 0.160$, (c) $\phi = 0.292$, and (d) $\phi = 0.532$.
From these snapshots, we can qualitatively understand the compressive behavior.
In the early stage (Panels (a)--(c)), the packing structure of small aggregates changes with minimal deformation of aggregates, while in the late stage (Panel (d)), deformation and compression of the aggregates become apparent.
We quantitatively discuss the temporal evolution of the packing structure and aggregate deformation in the following sections.
Figures for $r_{\rm agg} = 64 r_{\bullet}$ are provided in Supporting Information (Figure S1).

\begin{figure}
\centering
\includegraphics[width = \columnwidth]{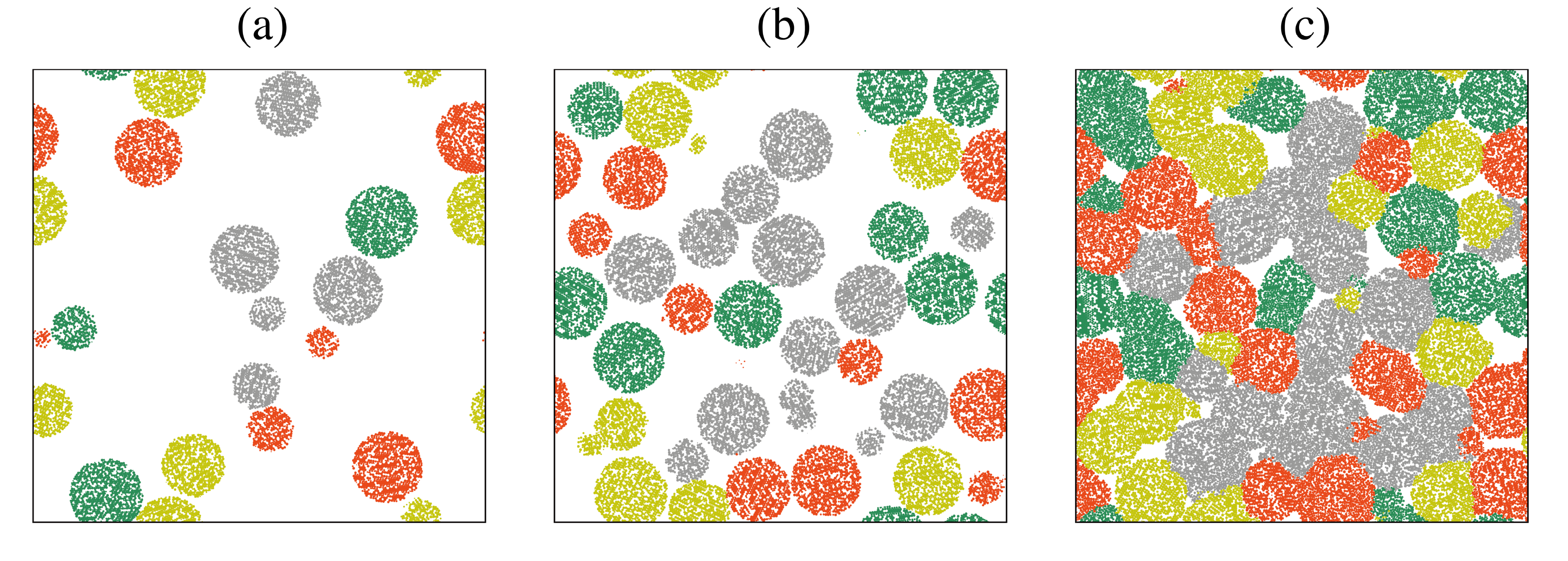}
\caption{
Cross-sections of the hierarchical granular pile with $r_{\rm agg} = 32 r_{\bullet}$ at a height of $z = L_{\rm z} / 2$.
(a) $\phi = 0.160$ (Figure \ref{fig:snapshot}(b)).
(b) $\phi = 0.292$ (Figure \ref{fig:snapshot}(c)).
(c) $\phi = 0.532$ (Figure \ref{fig:snapshot}(d)).
}
\label{fig:cut_32}
\end{figure}

Figure \ref{fig:cut_32} shows the cross-sections of the hierarchical granular pile from Figure \ref{fig:snapshot} at a height of $z = L_{\rm z} / 2$.
For $\phi = 0.160$ and $\phi = 0.292$ (Panels (a) and (b)), the cross sections of small aggregates are circular and retain their initial shape, whereas for $\phi = 0.532$ (Panel (c)), the aggregates undergo significant deformation.
By comparing the cases of $\phi = 0.292$ and $\phi = 0.160$, we observe that both the number and area of aggregate--aggregate contacts increase due to compression.
We investigate this change more quantitatively in Section \ref{sec:inter-aggregate}.

\begin{figure}
\centering
\includegraphics[width = \columnwidth]{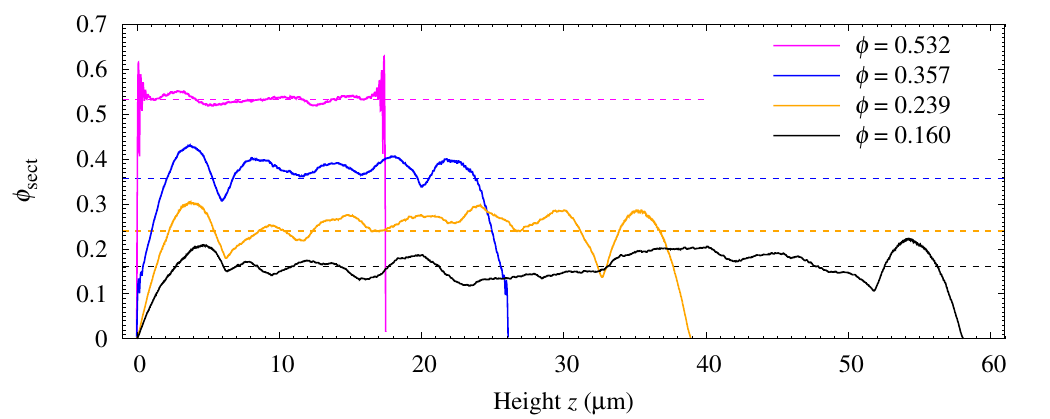}
\caption{
Vertical distribution of $\phi_{\rm sect}$ for $r_{\rm agg} = 32 r_{\bullet}$ (where $z = 0$ is at the bottom wall).
Different lines represent the distribution of $\phi_{\rm sect}$ at different values of $\phi$.
}
\label{fig:phi_height_32}
\end{figure}

We can define the filling factor at the height $z$, $\phi_{\rm sect} {( z )}$, as the cross-section at $z$, $\sigma_{\rm sect} {( z )}$, divided by the area of the simulation box, $L_{\rm x} L_{\rm y}$.
Figure \ref{fig:phi_height_32} shows the vertical distribution of $\phi_{\rm sect}$ for $r_{\rm agg} = 32 r_{\bullet}$.
We find that $\phi_{\rm sect}$ is approximately homogeneous, except in regions close to the top and bottom walls.
This homogeneity supports the conclusion that our numerical simulations correspond to quasi-static compression tests \citep{2012A&A...541A..59S}.
We also check the coupling of the pressure at the top and bottom walls in Section \ref{sec:pressure}.
In Figure \ref{fig:phi_height_32}, we observe a wavy pattern with a peak-to-peak distance of $\sim r_{\rm agg}$, although its amplitude is significantly smaller than $\phi$.
Both the peak-to-peak distance and amplitude of the wavy pattern vary due to the disordered nature of the aggregate packing structure (see Figures \ref{fig:snapshot} and \ref{fig:cut_32}).

For $\phi = 0.160$ and $\phi = 0.239$, $\phi_{\rm sect} \approx 0$ at the walls and it increases linearly with the distance from the walls.
This behavior is consistent with the observation that aggregates are in contact with walls without deformation.
In contrast, for $\phi = 0.357$, aggregates deform and form flat aggregate--wall contacts, which corresponds to a sharp increase in $\phi_{\rm sect}$ near the walls with a length scale of $\sim r_{\bullet}$.

As shown in Figures \ref{fig:snapshot}(d) and \ref{fig:cut_32}(c), aggregates are completely deformed at $\phi = 0.532$.
The vertical distribution of $\phi_{\rm sect}$ for $\phi = 0.532$ is homogeneous, except in regions close to the walls.
The length scale of the oscillation of $\phi_{\rm sect}$ near the walls is $\sim r_{\bullet}$, which might originate from the initial lattice-like configuration of constituent particles in small aggregates formed via the RBD process.

\subsection{Pressure at the top and bottom walls}
\label{sec:pressure}

\begin{figure}
\centering
\includegraphics[width = \columnwidth]{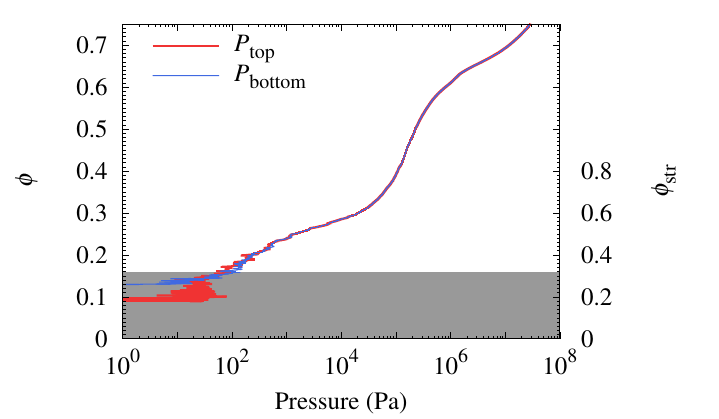}
\caption{
Pressures at the top and bottom walls, $P_{\rm top}$ and $P_{\rm bottom}$, with $r_{\rm agg} = 32 r_{\bullet}$.
In our simulation, the dynamical coupling between the top and bottom walls is not achieved when $\phi < 0.16$ (gray area).
We define the filling factor of the aggregate packing structure as $\phi_{\rm str} = \phi / \phi_{\rm agg}$.
}
\label{fig:P_phi_32}
\end{figure}

\begin{figure}
\centering
\includegraphics[width = \columnwidth]{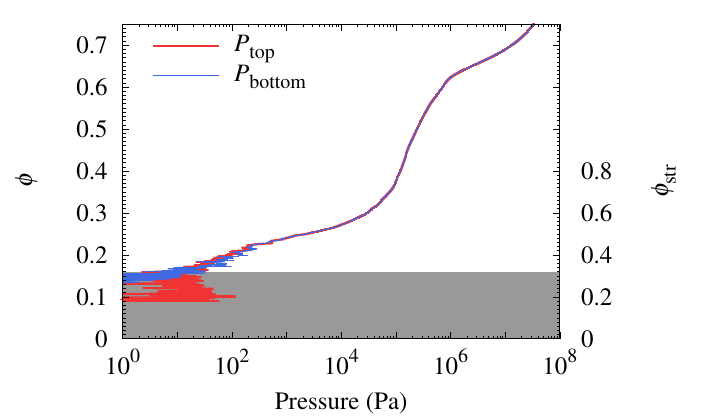}
\caption{
Same as Figure \ref{fig:P_phi_32}, but for $r_{\rm agg} = 64 r_{\bullet}$.
}
\label{fig:P_phi_64}
\end{figure}

Figures \ref{fig:P_phi_32} and \ref{fig:P_phi_64} represent the pressure at the top and bottom walls during compression, $P_{\rm top}$ and $P_{\rm bottom}$, with $r_{\rm agg} = 32 r_{\bullet}$ and $r_{\rm agg} = 64 r_{\bullet}$, respectively.
We find that $P_{\rm top} \approx P_{\rm bottom}$ if $\phi \ge 0.16$, whereas $P_{\rm top} \ne P_{\rm bottom}$ if $\phi < 0.16$.
In our simulation setup, small aggregates are not in contact with each other at $t = 0$.
The top wall moves with the velocity $v_{\rm wall}$ while the position of the bottom wall is fixed.
Therefore, there is a time lag for the initiation of particle--wall contacts between the top and bottom walls, and a finite time is needed to achieve the dynamical coupling of top and bottom walls.

As shown in Figures \ref{fig:P_phi_32} and \ref{fig:P_phi_64}, the compression curve of hierarchical granular piles has a winding shape, which corresponds to the fact that structure evolves through a multi-step process (see also Figure \ref{fig:schematic_curve}).
Laboratory experiments of the compression of hierarchical granular piles have also reported on the multi-step behavior \citep{2013Icar..226..111M, 2015Icar..257...33S}. 
We derive the semi-analytical model of the compression curve in Section \ref{sec:semi-analytic}.

We note that the compression curve is nearly independent of the setting of $v_{\rm wall}$, as long as the compression speed is sufficiently low to ensure quasi-static compression.
To verify this, we conduct an additional compression test using a constant $v_{\rm wall}$.
Specifically, we set $v_{\rm wall} = 0.31~{\rm m}~{\rm s}^{-1}$ and investigate the compression curve for $r_{\rm agg} = 32 r_{\bullet}$.
We confirm that the resulting compression curve is nearly identical to that obtained with a constant $\tau_{\rm comp}$ setting (see Figure S2 in Supporting Information).

\subsection{Inter-aggregate contacts within hierarchical granular piles}
\label{sec:inter-aggregate}

In this section, we show the evolution of the aggregate packing structure.
Here, we introduce the filling factor of the aggregate packing structure, $\phi_{\rm str}$.
When aggregates retain their initial shape in a hierarchical granular pile, $\phi = \phi_{\rm agg} \phi_{\rm str}$, with $\phi_{\rm agg} = 0.5$.
In other words, $\phi_{\rm str}$ is given by $\phi_{\rm str} = \phi / \phi_{\rm agg}$.

We evaluate the aggregate--aggregate contact properties by the following procedure.
First, we calculate the center of mass of each aggregate.
Then, we determine the inter-aggregate distance between the centers of mass, $d_{\rm agg}$, for each aggregate--aggregate pair.
Finally, we identify the aggregate pairs in contact (i.e., $d_{\rm agg} \le 2 r_{\rm agg}$) and obtain the aggregate--aggregate compression length, $\delta_{\rm agg} = 2 r_{\rm agg} - d_{\rm agg}$, for each pair.
We repeat the procedure above for each snapshot.

\begin{figure}
\centering
\includegraphics[width = 0.8\columnwidth]{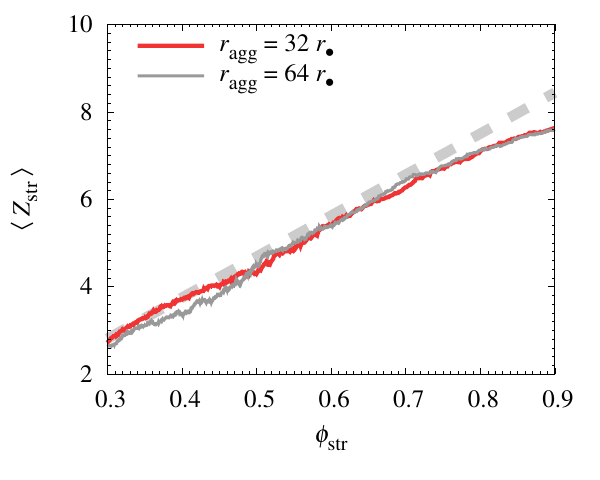}
\caption{
Average coordination number for aggregate--aggregate contacts, ${\langle Z_{\rm str} \rangle}$, as a function of $\phi_{\rm str}$.
The red and gray lines correspond to $r_{\rm agg} = 32 r_{\bullet}$ and $64 r_{\bullet}$, respectively.
The gray dashed line represents an empirical fit: ${\langle Z_{\rm str} \rangle} = {( 6 / 0.64 )} \phi_{\rm str}$ (Equation (\ref{eq:Z_str_phi_str})).
}
\label{fig:Z_phi_32}
\end{figure}

\begin{figure}
\centering
\includegraphics[width = 0.8\columnwidth]{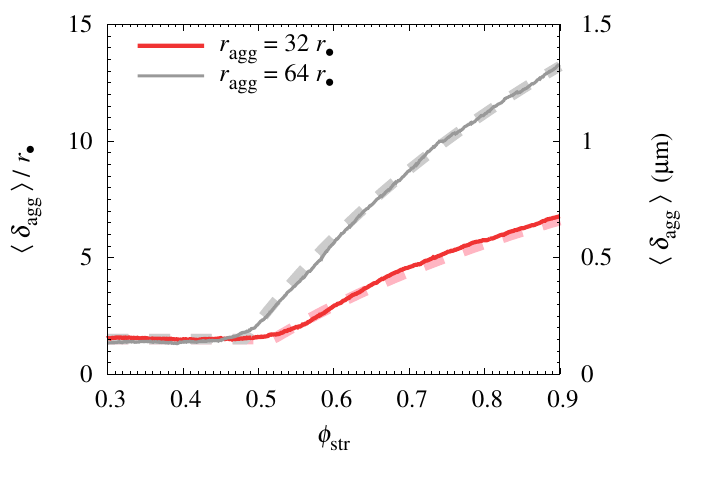}
\caption{
Average of the compression length for aggregate--aggregate contacts, ${\langle \delta_{\rm agg} \rangle}$, as a function of $\phi_{\rm str}$.
The red and gray lines correspond to $r_{\rm agg} = 32 r_{\bullet}$ and $64 r_{\bullet}$, respectively.
The dashed lines represent empirical fits (Equation (\ref{eq:delta_agg_phi_str})).
}
\label{fig:delta_phi_32}
\end{figure}

Figure \ref{fig:Z_phi_32} shows the average coordination number for aggregate--aggregate contacts, ${\langle Z_{\rm str} \rangle}$, with respect to $\phi_{\rm str}$.
Here, ${\langle Z_{\rm str} \rangle}$ is given by ${\langle Z_{\rm str} \rangle} = 2 n_{\rm c} / N_{\rm agg}$, where $n_{\rm c}$ is the number of aggregate--aggregate pairs in contact.
For $0.3 < \phi_{\rm str} < 0.7$, we find that ${\langle Z_{\rm str} \rangle}$ is approximately given by
\begin{equation}
{\langle Z_{\rm str} \rangle} = {( 6 / 0.64 )} \phi_{\rm str},
\label{eq:Z_str_phi_str}
\end{equation}
and it is independent of $r_{\rm agg}$.
Figure \ref{fig:delta_phi_32} shows the average of the compression length for aggregate--aggregate contacts, ${\langle \delta_{\rm agg} \rangle}$, with respect to $\phi_{\rm str}$.
In both cases where $r_{\rm agg} = 32 r_{\bullet}$ and $64 r_{\bullet}$, ${\langle \delta_{\rm agg} \rangle} \approx 1.5 r_{\bullet}$ if $\phi_{\rm str}$ is below a certain threshold, and it is independent of $r_{\rm agg}$.
In contrast, when $\phi_{\rm str}$ exceeds this threshold, ${\langle \delta_{\rm agg} \rangle}$ is proportional to $r_{\rm agg}$ and increases with $\phi_{\rm str}$.
We find that the following equation well reproduces the numerical results:
\begin{equation}
{\langle \delta_{\rm agg} \rangle} = \max{\left[ \delta_{\rm agg, 0}, \delta_{\rm agg, comp} \right]},
\label{eq:delta_agg_phi_str}
\end{equation}
where $\delta_{\rm agg, 0} = 1.5 r_{\bullet} = 0.15~{\mu}{\rm m}$ and
\begin{equation}
\delta_{\rm agg, comp} = \alpha_{\rm agg} \cdot 2 r_{\rm agg} {\left[ 1 - {\left( \frac{\phi_{\rm str}}{0.45} \right)}^{- 1/3} \right]},
\label{eq:delta_agg_comp}
\end{equation}
with a fitting coefficient $\alpha_{\rm agg} = 0.5$.
The threshold $\phi_{\rm str}$ value can be obtained by solving the equation $\delta_{\rm agg, comp} = \delta_{\rm agg, 0}$, and it approaches $0.45$ as $r_{\rm agg} / r_{\bullet}$ becomes infinitely large.

\begin{figure}
\centering
\includegraphics[width = 0.8\columnwidth]{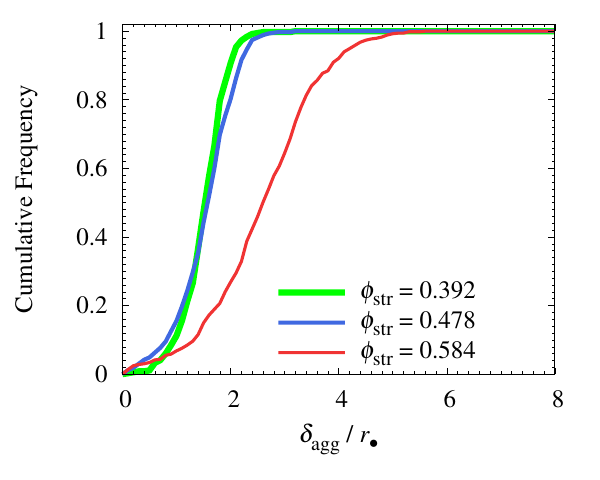}
\caption{
Cumulative frequency distribution of $\delta_{\rm agg}$ for $r_{\rm agg} = 32 r_{\bullet}$.
The different lines represent the distributions at different values of $\phi_{\rm str}$.
}
\label{fig:delta_32_step}
\end{figure}

We also calculate the cumulative frequency distribution of $\delta_{\rm agg}$ for $r_{\rm agg} = 32 r_{\bullet}$ (Figure \ref{fig:delta_32_step}).
For $\phi_{\rm str} = 0.392$ and $0.478$, approximately 80\% of all aggregate--aggregate contacts have a compression length in the narrow range of $1 < \delta_{\rm agg} / r_{\bullet} < 2$.
In contrast, for $\phi_{\rm str} = 0.584$, the $\delta_{\rm agg} / r_{\bullet}$ values are broadly distributed in the range of $1 < \delta_{\rm agg} / r_{\bullet} < 4$ for 80\% of all aggregate--aggregate contacts.

\begin{figure}
\centering
\includegraphics[width = 0.8\columnwidth]{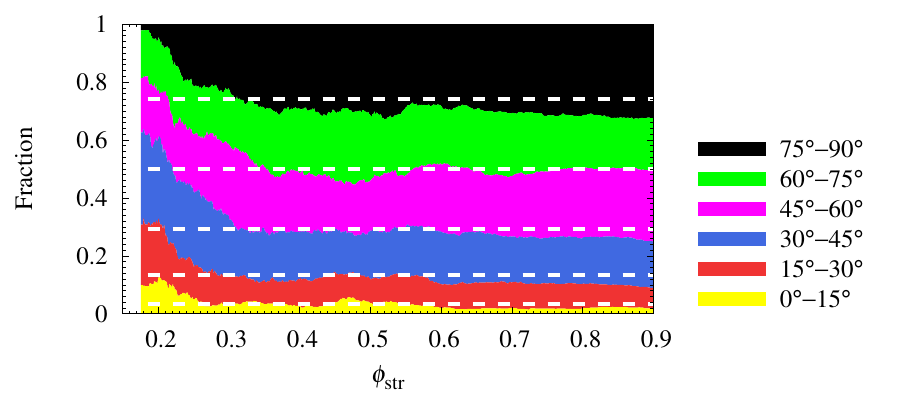}
\caption{
Distribution of the directions of aggregate--aggregate contacts as a function of $\phi_{\rm str}$.
The angle is defined as $0^{\circ}$ when the direction of the aggregate--aggregate contact is parallel to the z-axis.
The white dashed lines correspond to the prediction for the homogeneous case.
}
\label{fig:angle_phi_32}
\end{figure}

We prepare the initial setting of the aggregate packing structure using the RBD process.
In the RBD process, spheres are deposited from one direction (see Figure \ref{fig:initial_schematic}), and the resulting structure is non-isotropic \citep{2014M&PS...49..109S}.
Figure \ref{fig:angle_phi_32} shows the distribution of the directions of aggregate--aggregate contacts.
Here, we define the angle as $0^{\circ}$ when the direction of the aggregate--aggregate contact is parallel to the z-axis.
If the distribution of the directions is homogeneous over $4\pi$ steradians, the fraction of contacts with angles between $\theta_{1}$ and $\theta_{2}$ ($0^{\circ} < \theta_{1} < \theta_{2} < 90^{\circ}$) is given by $\sin{\theta_{2}} - \sin{\theta_{1}}$.
The white dashed lines correspond to the prediction for the homogeneous case.
For $\phi_{\rm str} < 0.35$, the fraction of contacts with small angles is clearly higher than expected for the homogeneous case.
In contrast, for $\phi_{\rm str} > 0.35$, the distribution is roughly consistent with the homogeneous case especially when we focus on the fractions of contacts with angles smaller and larger than $60^{\circ}$ (50\% and 50\%; the border between magenta and green).

\subsection{Rotational and translational motions of aggregates}

As shown in Figure \ref{fig:snapshot}, in the first stage, the compression of hierarchical granular piles is accompanied by the rearrangement of the aggregate packing structure.
In the second stage, the compression of hierarchical granular piles is predominantly due to the deformation of the small aggregates.
In this section, we present the rotational and translational velocities of the aggregates within the hierarchical granular piles.

We define ${\bm x}_{\rm agg} = {( x_{\rm agg}, y_{\rm agg}, z_{\rm agg} )}$ as the position of the center of mass of the aggregate.
The translational velocity of the aggregate is given by $\dot{\bm x}_{\rm agg}$.
The velocity of the top wall is ${\bm v}_{\rm wall} = {( 0, 0, - v_{\rm wall} )}$.
Thus, if the computational domain shrinks homogeneously, the reference velocity at ${\bm x}_{\rm agg}$ is ${\bm v}_{\rm ref} = {( 0, 0, {- z_{\rm agg} v_{\rm wall} / L_{\rm z}} )}$.
The magnitude of the relative translational velocity of the aggregate, $V_{\rm tra}$, is given by the following equation:
\begin{equation}
V_{\rm tra} = { || \dot{\bm x}_{\rm agg} - {\bm v}_{\rm ref} || }.
\end{equation}

Similarly, we define the position of the $i$th particle within the aggregate as ${\bm x}_{i} = {( x_{i}, y_{i}, z_{i} )}$ and its velocity as $\dot{\bm x}_{i}$.
Assuming the aggregate follows a rigid body motion, we can define $\dot{\bm x}_{i, {\rm rot}} = \dot{\bm x}_{i} - \dot{\bm x}_{\rm agg}$ as the rotational velocity of the $i$th particle.
The rotational energy of the aggregate is $K_{\rm rot} = \sum_{i} {( m_{\bullet} {( || \dot{\bm x}_{i, {\rm rot}} || )}^{2} / 2 )}$, and the moment of inertia is $I_{\rm agg} = {( 2 / 5 )} m_{\bullet} N_{\rm par} {r_{\rm agg}}^{2}$.
Using these values, we can evaluate the magnitude of the rotational velocity of the aggregate, $V_{\rm tra}$, as follows:
\begin{equation}
V_{\rm rot} = \sqrt{\frac{K_{\rm rot}}{{( 1 / 2 )} I_{\rm agg}}} r_{\rm agg}.
\end{equation}

\begin{figure}
\centering
\includegraphics[width = \columnwidth]{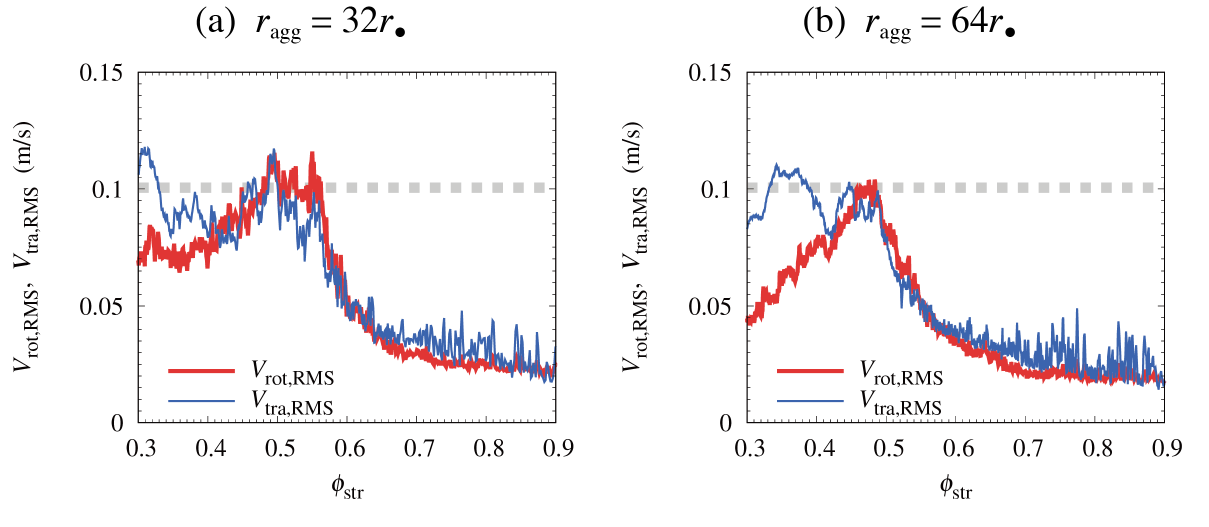}
\caption{
Root mean squares of the rotational and relative translational velocities of aggregates, $V_{\rm rot, RMS}$ and $V_{\rm tra, RMS}$, respectively.
(a) $r_{\rm agg} = 32 r_{\bullet}$.
(b) $r_{\rm agg} = 64 r_{\bullet}$.
The gray dotted lines represent the characteristic velocity, $V_{\rm ch} = {( 2 \pi r_{\rm agg} )} / \tau_{\rm comp}$ (Equation (\ref{eq:V_ch})).
}
\label{fig:v_phi}
\end{figure}

Figure \ref{fig:v_phi} shows the root mean squares of $V_{\rm rot}$ and $V_{\rm tra}$ within hierarchical granular piles, denoted as $V_{\rm rot, RMS}$ and $V_{\rm tra, RMS}$.
We find that $V_{\rm rot, RMS} \approx V_{\rm tra, RMS}$ at $\phi_{\rm str} \ge 0.4$.
Both $V_{\rm rot, RMS}$ and $V_{\rm tra, RMS}$ sharply drop at $\phi_{\rm str} = 0.55$ for $r_{\rm agg} = 32 r_{\bullet}$ and at $\phi_{\rm str} = 0.48$ for $r_{\rm agg} = 64 r_{\bullet}$.
The values of $\phi_{\rm str}$ at the transition correspond to the onset of an increase in the compression length for aggregate--aggregate contacts (see Figure \ref{fig:delta_phi_32}).
These results support that the rearrangement of the aggregate packing structure occurs in the first stage of compression.
Our findings also highlight the multi-step nature of the compression of hierarchical granular piles and the changes in the structural evolution mechanisms.

Assuming that the timescale for the rotational and translational motions of aggregates is comparable to the compression timescale ($= \tau_{\rm comp}$), the characteristic velocity, $V_{\rm ch}$, is approximately given by
\begin{equation}
V_{\rm ch} = \frac{2 \pi r_{\rm agg}}{\tau_{\rm comp}}.
\label{eq:V_ch}
\end{equation}
The gray dotted lines in Figure \ref{fig:v_phi} denotes $V_{\rm ch}$.
It appears that $V_{\rm rot, RMS} \approx V_{\rm tra, RMS} \approx V_{\rm ch}$ near the transition point for the velocity trends.

\section{Compression test of single aggregate}
\label{sec:single}

In Section \ref{sec:results}, we revealed that the deformation of small aggregates within hierarchical granular piles is the key process in the second stage where $\phi_{\rm str} > 0.5$.
In this section, we perform compression tests on a single aggregate using the same DEM model as in Section \ref{sec:results}, which enable us to construct a mechanical model for the deformation of each aggregate.
We find that the stress--strain relationship of the aggregates is well approximated by that of plastic spheres.

\begin{figure}
\centering
\includegraphics[width = \columnwidth]{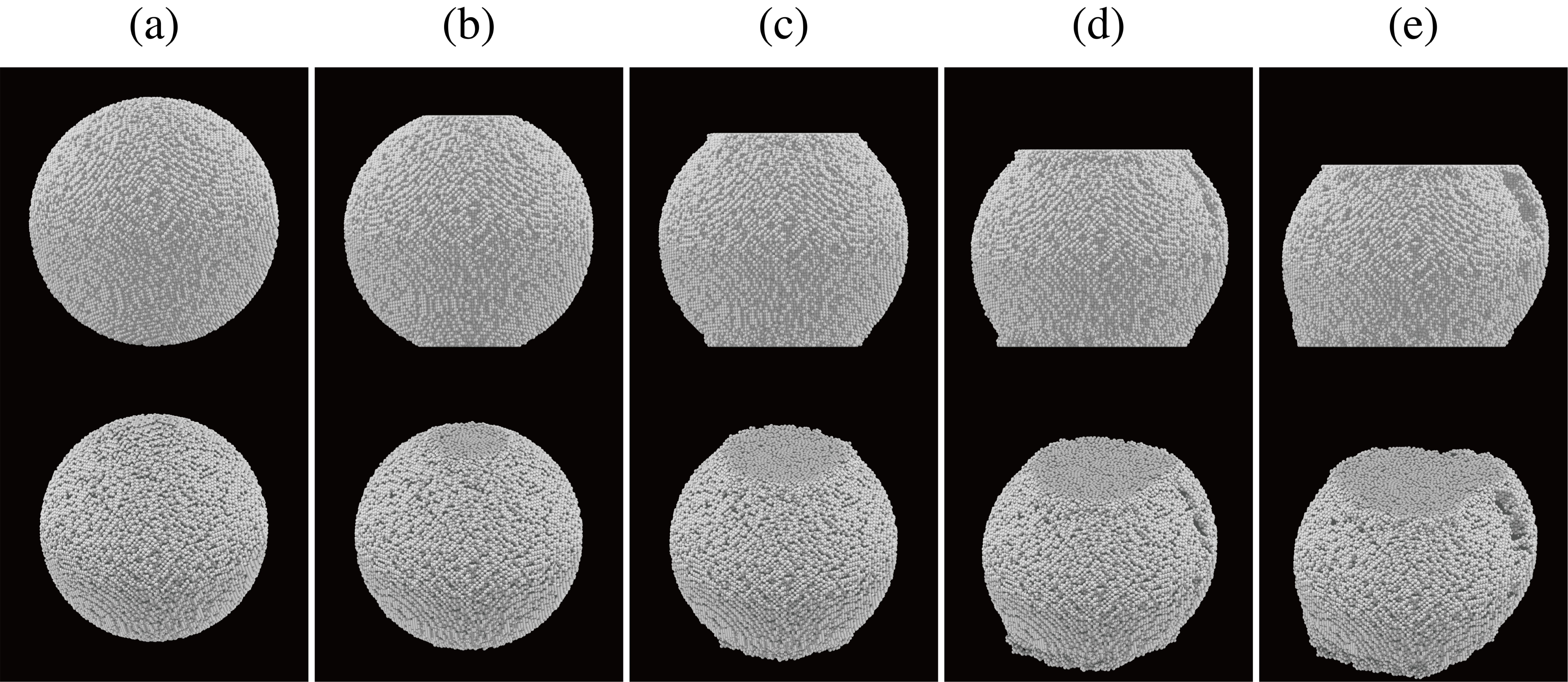}
\caption{
Snapshots of a compression test of a single aggregate (Run \#1 for $r_{\rm agg} = 64 r_{\bullet}$).
(a) Initial condition.
(b) $\delta_{\rm wall} = 0.98~{\mu}{\rm m}$. 
(c) $\delta_{\rm wall} = 1.89~{\mu}{\rm m}$. 
(d) $\delta_{\rm wall} = 2.73~{\mu}{\rm m}$. 
(e) $\delta_{\rm wall} = 3.51~{\mu}{\rm m}$. 
The upper panels show edge-on views, and the lower ones are taken from an oblique upper angle.
}
\label{fig:snapshot_single}
\end{figure}

We perform five simulation runs for each aggregate size.
The compression timescale is set to $\tau_{\rm comp} = 1 \times 10^{-4}~{\rm s}$ for both $r_{\rm agg} = 32 r_{\bullet}$ and $64 r_{\bullet}$.
Figure \ref{fig:snapshot_single} shows snapshots of a compression test of a single aggregate with $r_{\rm agg} = 64 r_{\bullet}$.
We define $\delta_{\rm wall} = 2 r_{\rm agg} - L_{\rm z}$ as the compression length.
We find that circular contact areas form at the top and bottom for small $\delta_{\rm wall}$ values (Panels (b) and (c)).
We also find that a fault-like feature forms when $\delta_{\rm wall}$ becomes large (Panels (d) and (e)).

\begin{figure}
\centering
\includegraphics[width = \columnwidth]{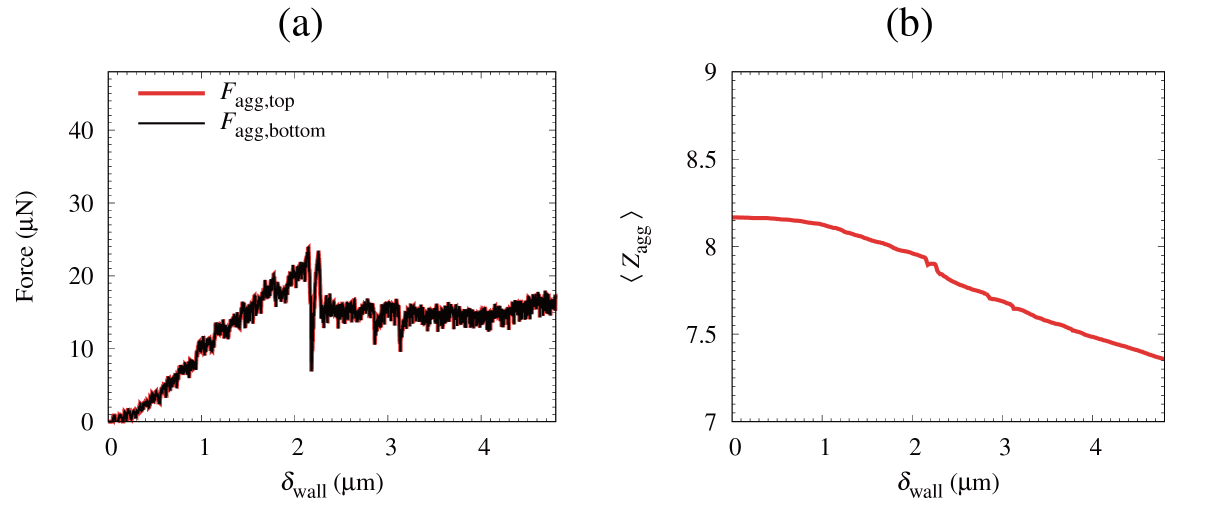}
\caption{
Force at the walls and the average coordination number.
(a) Force--displacement relationship for a compression test of a single aggregate.
(b) Average coordination number as a function of $\delta_{\rm wall}$.
The results are presented for the run shown in Figure \ref{fig:snapshot_single}.
}
\label{fig:force_single_64_case1}
\end{figure}

Figure \ref{fig:force_single_64_case1}(a) shows the force--displacement relationship for the run shown in Figure \ref{fig:snapshot_single}.
We find that the forces at the top and bottom walls, $F_{\rm agg, top}$ and $F_{\rm agg, bottom}$, are approximately the same, indicating that the compression speed is sufficiently small and the dynamical coupling of top and bottom walls is achieved.
For $\delta_{\rm wall} < 2~{\mu}{\rm m}$, the forces increase approximately linearly with $\delta_{\rm wall}$.
In contrast, for $\delta_{\rm wall} > 2~{\mu}{\rm m}$, the forces are approximately constant.
The transition roughly corresponds to the formation of a fault-like feature (see Figure \ref{fig:snapshot_single}).
The force--displacement relationship for the other four runs are provided in Supporting Information (Figure S3).

\begin{figure}
\centering
\includegraphics[width = \columnwidth]{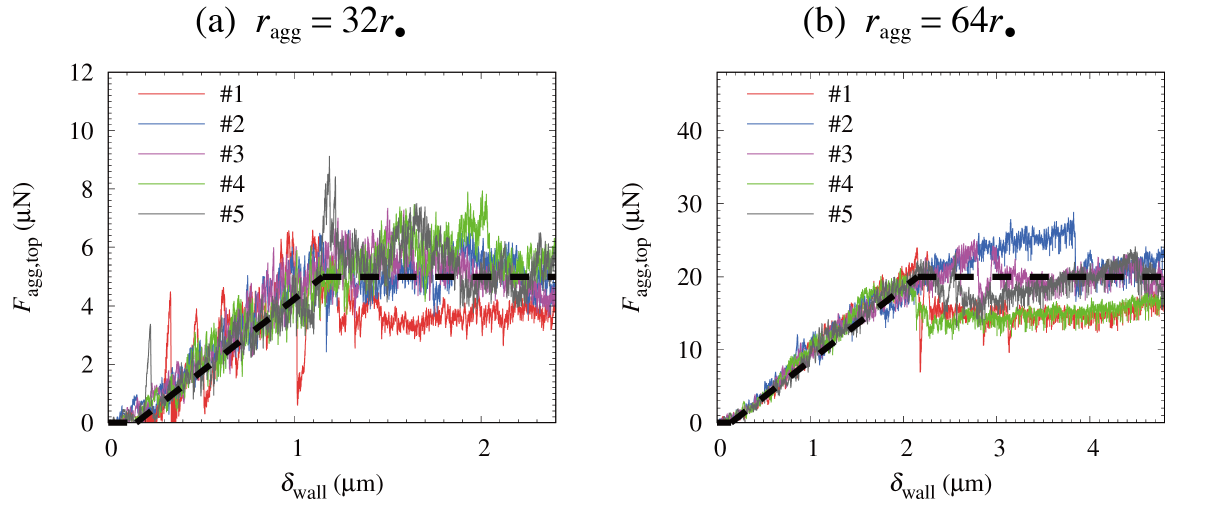}
\caption{
Force--displacement relationship for single aggregates.
We performed five simulation runs for both the $r_{\rm agg} = 32 r_{\bullet}$ and $64 r_{\bullet}$ cases.
(a) $r_{\rm agg} = 32 r_{\bullet}$.
(b) $r_{\rm agg} = 64 r_{\bullet}$.
Black dashed lines represent an semi-analytical fit (Equation (\ref{eq:F_agg_wall})).
}
\label{fig:force_single}
\end{figure}

Figure \ref{fig:force_single} shows the force--displacement relationship for the five runs for each aggregate size.
We find that the following equation can reproduce the general trend of the force--displacement relationship (black dashed lines):
\begin{equation}
F_{\rm agg, top} = \left\{ 
\begin{array}{ll}
0,   &   ( \delta_{\rm wall} < \delta_{\rm offset} )  \\
k_{\rm agg} {\left( \delta_{\rm wall} - \delta_{\rm offset} \right)}, & ( \delta_{\rm offset} \le \delta_{\rm wall} < \delta_{\rm crit} ) \\
F_{\rm max, agg},   & ( \delta_{\rm wall} \ge \delta_{\rm crit} )
\end{array}
\right.
\label{eq:F_agg_wall}
\end{equation}
where $\delta_{\rm offset}$ is the offset, $k_{\rm agg} = k_{32} {\left[ r_{\rm agg} / {( 32 r_{\bullet} )} \right]}$ is the proportionality constant, and $F_{\rm max, agg} = F_{32} {\left[ r_{\rm agg} / {( 32 r_{\bullet} )} \right]}^{2}$ is the maximum force.
The critical compression length, $\delta_{\rm crit}$, is given by $\delta_{\rm crit} = F_{\rm max, agg} / k_{\rm agg} + \delta_{\rm offset}$.
We set $\delta_{\rm offset} = 1.5 r_{\bullet} = 0.15~{\mu}{\rm m}$, which is equal to the value of $\delta_{\rm agg, 0}$ in Equation (\ref{eq:delta_agg_phi_str}).
Two fitting constants, $k_{32}$ and $F_{32}$, are chosen to align with numerical results and are set to $k_{32} = 5~{\rm N}~{\rm m}^{-1}$ and $F_{32} = 5~{\mu}{\rm N}$.

Here, we provide a theoretical interpretation of Equation (\ref{eq:F_agg_wall}).
If the aggregates behave as perfectly plastic spheres with a yield stress of $Y_{\rm agg}$, the force at the wall should be approximately given by $F_{\rm agg, top} \approx {( \pi / 2 )} Y_{\rm agg} r_{\rm agg} \delta_{\rm wall}$ \citep{doi:10.1080/14786443008565033, 2024arXiv240815573A}.
Considering that there is an offset between the apparent and effective surfaces for the aggregate--wall contacts, we can modify the equation above as follows: $F_{\rm agg, top} \approx {( \pi / 2 )} Y_{\rm agg} r_{\rm agg} {( \delta_{\rm wall} - \delta_{\rm offset} )}$.
Then, we can estimate of the yield stress of aggregates:
\begin{equation}
Y_{\rm agg} \approx \frac{2 k_{\rm agg}}{\pi r_{\rm agg}} = 1.0~{\rm MPa}.
\end{equation}
It should be noted that $Y_{\rm agg}$ is independent of $r_{\rm agg}$.

The aggregates used in this study are prepared by the CPE process, and their filling factor is $\phi_{\rm agg} = 0.5$.
Figure \ref{fig:force_single_64_case1}(b) shows the average coordination number, ${\langle Z_{\rm agg} \rangle}$, which is ${\langle Z_{\rm agg} \rangle} \approx 8$ at the initial condition, $\delta_{\rm wall} = 0$.
As ${\langle Z_{\rm agg} \rangle}$ within the aggregate is high, the breaking of inter-particle contact is required to cause plastic deformation.
The decreasing trend of ${\langle Z_{\rm agg} \rangle}$ shown in Figure \ref{fig:force_single_64_case1}(b) supports this idea.
The corresponding stress would be roughly given by $\sim {( F_{\rm crit} )} / {( \pi {r_{\bullet}}^{2} )} = 1.5~{\rm MPa}$, and this order-of-magnitude estimate explains the $Y_{\rm agg}$ value obtained from numerical simulations.

\section{Semi-analytic formula for the compression curve}
\label{sec:semi-analytic}

We obtained the compression curve for hierarchical granular piles using DEM simulations with millions of particles in Section \ref{sec:results}.
We also derived the force--displacement relationship for small spherical aggregates in Section \ref{sec:single}.
In this section, we propose a semi-analytic formula for the compression curve of hierarchical granular piles.
We find that the compression curve can be divided into three stages: (i) rearrangement of the aggregate packing structure, (ii) plastic deformation of small aggregates, and (iii) elastic deformation of constituent particles (see Figures \ref{fig:schematic_curve} and \ref{fig:semi-analytic}).

\subsection{Stage 1: rearrangement of the aggregate packing structure}
\label{sec:stage1}

At the early stage of compression, where $\phi_{\rm str}$ is sufficiently small, the structural change of hierarchical granular piles occurs without deformation of the aggregates.
The yield strength of hierarchical granular piles would be proportional to the tangential friction between aggregates \citep{2013A&A...554A...4K}.
Note that we do not consider tangential friction in particle--particle interactions in this study (see Section \ref{sec:DEM}).

The blue curves in Figure \ref{fig:semi-analytic} denote the semi-analytic equation of the compressive curve for Stage 1.
We fit the compressive curve for Stage 1 using a modified polytropic equation \citep{2009ApJ...701..130G, 2012A&A...541A..59S}:
\begin{equation}
P_{\rm S1} = P_{\rm mid} {\left( \frac{\phi_{\rm str} - \phi_{\rm min}}{\phi_{\rm max} - \phi_{\rm str}} \right)}^{1 + 1/n},
\label{eq:P_S1}
\end{equation}
where $P_{\rm S1}$ is the pressure in Stage 1, $\phi_{\rm min} = 0.15$ \citep{2004PhRvL..93k5503B} and $\phi_{\rm max} = 0.64$ \citep{1983PhRvA..27.1053B} are the minimum and maximum values of $\phi_{\rm str}$.
The polytropic index is set to $n = 1$, although it is difficult to precisely determine the index due to the large scatter in our numerical results.
We also obtain the prefactor $P_{\rm mid}$ from numerical results as follows:
\begin{equation}
P_{\rm mid} = 240 {\left( \frac{r_{\rm agg}}{32 r_{\bullet}} \right)}^{- 3/2}~{\rm Pa}.
\label{eq:P_mid}
\end{equation}
Based on a theoretical investigation, we find that $P_{\rm mid}$ should be proportional to ${r_{\rm agg}}^{- 3/2}$ when the friction for aggregate--aggregate rotation plays a key role in Stage 1 (see the following paragraphs).

To support the proposed model, we outline its underlying basis.
\cite{2013A&A...554A...4K} have performed DEM simulations of the slow compression of fractal aggregates consisting of frictional spheres.
For highly porous aggregates, they found that the filling factor dependence of compressive pressure is approximately given by the polytropic equation:
\begin{equation}
P_{\rm K13} \approx P_{\rm rot} \phi^{1 + 1/n},
\label{eq:K13}
\end{equation}
where the prefactor $P_{\rm rot}$ is approximately equal to the work needed to roll a sphere against its friction divided by the volume of the sphere.
The index $n$ would depend on the fractal structure of the initial aggregates \citep{2019PTEP.2019i3E02A, 2024JPSJ...93b4401A}.
Although Equation (\ref{eq:K13}) was originally proposed for the compression curve of fractal aggregates consisting of frictional spheres, it is also applicable to the compression curve of hierarchical granular piles.
We redefine $P_{\rm rot}$ as the the work needed to roll an aggregate against its friction divided by the volume of the aggregate.
The compressive pressure is given by
\begin{equation}
P_{\rm K13} \approx P_{\rm rot} {\phi_{\rm str}}^{1 + 1/n}.
\label{eq:K13_agg}
\end{equation}
Equation (\ref{eq:K13_agg}) provides a good approximation to Equation (\ref{eq:P_S1}) when the following condition, $\phi_{\rm min} \ll \phi_{\rm str} \ll \phi_{\rm max}$, is satisfied.

We evaluate the value of $P_{\rm rot}$ for small aggregates used in our simulations by an order-of-magnitude estimate.
For two aggregates in contact, the radius of the inter-aggregate contact area, $a_{\rm agg}$, is expressed as a function of their compression length, $\delta_{\rm agg}$, and the aggregate radius, $r_{\rm agg}$.
When the deformation of the aggregates is negligible, $\delta_{\rm agg}$ is given by $\delta_{\rm agg} \approx \delta_{\rm agg, 0}$, and $a_{\rm agg}$ is approximated by $a_{\rm agg} \approx \sqrt{r_{\rm agg} \delta_{\rm agg, 0}} \approx \sqrt{1.5 r_{\rm agg} r_{\bullet}}$.
Here, we consider a situation where one aggregate rotates around another while maintaining their contact without sliding.
The length of the great circle of an aggregate is $2 \pi r_{\rm agg}$, and the area where contact is made by one aggregate--aggregate rotation, $S_{\rm rot}$, is given by $S_{\rm rot} \approx 2 \pi r_{\rm agg} \cdot 2 a_{\rm agg}$.
The number of particle--particle connection and disconnection events per one rotation, $N_{\rm rot}$, can be approximated as $N_{\rm rot} \sim {( S_{\rm rot} \phi_{\rm agg} )} / {( \pi {r_{\bullet}}^{2} )} \sim 2.4 {( r_{\rm agg} / r_{\bullet} )}^{3/2}$.
The work needed to break one particle--particle contact is $W_{\rm break} \approx 1.3 F_{\rm crit} \delta_{0}$ \citep{1971RSPSA.324..301J, 2007ApJ...661..320W}.
Therefore, we can roughly evaluate $P_{\rm rot}$ as follows:
\begin{equation}
P_{\rm rot} \approx \frac{N_{\rm rot} W_{\rm break}}{{( 4 \pi / 3 )} {r_{\rm agg}}^{3}} \approx 200 {\left( \frac{r_{\rm agg}}{32 r_{\bullet}} \right)}^{- 3/2}~{\rm Pa}.
\end{equation}
Our order-of-magnitude estimate reproduces the numerical results well (Equation (\ref{eq:P_mid})).

\subsection{Stage 2: plastic deformation of small aggregates}
\label{sec:stage2}

As the compression of hierarchical granular piles progresses, ${\langle Z_{\rm str} \rangle}$ increases, and each aggregate can no longer rotate freely.
The compression in Stage 2 is caused by the deformation of the aggregates, and repulsive forces act between the aggregates.

The relationship between the pressure in Stage 2, $P_{\rm S2}$, and the average of the interaggregate force, ${\langle F_{\rm agg} \rangle}$, is approximately given by the following equation \citep{https://doi.org/10.1002/cite.330420806, 2024PhRvE.109b4904A}:
\begin{equation}
P_{\rm S2} = c_{2} \frac{{\langle Z_{\rm str} \rangle} {\langle F_{\rm agg} \rangle}}{4 \pi {r_{\rm agg}}^{2}} \frac{2 r_{\rm agg} - {\langle \delta_{\rm agg} \rangle}}{2 r_{\rm agg}} \phi_{\rm str}.
\label{eq:P_S2}
\end{equation}
The dependence of ${\langle Z_{\rm str} \rangle}$ on $\phi_{\rm str}$ is given by Equation (\ref{eq:Z_str_phi_str}).
The dependence of ${\langle \delta_{\rm agg} \rangle}$ on $\phi_{\rm str}$ is given by Equation (\ref{eq:delta_agg_phi_str}).
Additionally, we assume that the relationship between ${\langle F_{\rm agg} \rangle}$ and ${\langle \delta_{\rm agg} \rangle}$ is identical to that for $F_{\rm agg, top}$ and $\delta_{\rm wall}$ (Equation (\ref{eq:F_agg_wall})).
We find that the equation above reproduces our numerical results well if we introduce a correction factor of $c_{2} = 1.2$ (green curves in Figure \ref{fig:semi-analytic}).

For the large aggregate limit (i.e., $r_{\rm agg} / r_{\bullet} \to \infty$), the transition from Stage 1 to Stage 2 occurs at $\phi_{\rm str} = 0.45$.
Assuming that the relationship between $\phi_{\rm str}$ and ${\langle Z_{\rm str} \rangle}$ is given by Equation (\ref{eq:Z_str_phi_str}), the transition occurs at ${\langle Z_{\rm str} \rangle} \approx 4$.
For random packings of monodisperse spheres with large sliding friction, several studies have shown that the interparticle normal force becomes positive when ${\langle Z_{\rm str} \rangle} \ge 4$ \citep{2008Natur.453..629S, 2020PhRvE.102c2903S}.
The transition observed in our numerical simulations corresponds to that reported in these previous studies.

\subsection{Stage 3: elastic deformation of constituent particles}

As shown in Figure \ref{fig:snapshot}(d), aggregates within hierarchical granular piles cannot retain their spherical shapes when $\phi_{\rm str} \ge 1$.
In our numerical simulation, the constituent particles are elastic nanospheres with surface energy.
For $\phi$ values higher than the jamming point (Stage 3), repulsive forces act between the constituent particles due to their elastic deformation.

As in Stage 2, the pressure in Stage 3, $P_{\rm S3}$, is approximately given by the following equation: 
\begin{equation}
P_{\rm S3} = c_{3} \frac{{\langle Z \rangle} {\langle F \rangle}}{4 \pi {r_{\bullet}}^{2}} \frac{2 r_{\bullet} - {\langle \delta \rangle}}{2 r_{\bullet}} \phi,
\label{eq:P_S3}
\end{equation}
where ${\langle Z \rangle}$ is the average coordination number for interparticle contacts within the hierarchical granular pile.
Figure \ref{fig:Z_phi_S3} shows the relationship between ${\langle Z \rangle}$ and $\phi$ in our simulations.
For $\phi \ge 0.64$, we find that ${\langle Z \rangle}$ is approximately given by
\begin{equation}
{\langle Z \rangle} = 20 {( \phi - \phi_{3} )},
\end{equation}
where $\phi_{3} = 0.3 - 0.02 {( r_{\rm agg} / {( 32 r_{\bullet} )} )}^{-1}$.
We also assume that the dependence of ${\langle \delta \rangle}$ on $\phi$ is given by the following equation:
\begin{equation}
{\langle \delta \rangle} - \delta_{0} = \alpha {\left( 2 r_{\bullet} - \delta_{0} \right)} {\left[ 1 - {\left( \frac{\phi}{0.64} \right)}^{- 1/3} \right]}.
\label{eq:delta_phi_S3}
\end{equation}
We set $\alpha = 0.5$, as in the case of $\alpha_{\rm agg}$ for ${\langle \delta_{\rm agg} \rangle}$ (see Equation (\ref{eq:delta_agg_comp})).
Note that ${\langle \delta \rangle} = \delta_{0}$ when no external forces act on the constituent particles (Figure \ref{fig:JKR}(b)).
Equation (\ref{eq:delta_phi_S3}) indicates that ${\langle \delta \rangle} > \delta_{0}$ for $\phi > 0.64$, with the critical $\phi$ value taken from the canonical jamming point for the packing of frictionless spheres \citep{1983PhRvA..27.1053B}.
The relationship between ${\langle F \rangle}$ and ${\langle \delta \rangle}$ is given by Equations (\ref{eq:JKR_a}) and (\ref{eq:JKR_F}).
We find that Equation (\ref{eq:P_S3}) reproduces our numerical results well if we introduce a correction factor of $c_{3} = 1.1$ (gray curves in Figure \ref{fig:semi-analytic}).

Figure \ref{fig:Z_phi_S3} also supports our findings that the compression of hierarchical granular piles proceeds in three stages.
For $r_{\rm agg} = 32 r_{\bullet}$, ${\langle Z \rangle}$ remains nearly constant for $\phi_{\rm str} \le 0.5$, which is consistent with the fact that the deformation of aggregates is negligible at Stage 1.
The ${\langle Z \rangle}$ value gradually decreases with increasing $\phi_{\rm str}$ for $\phi_{\rm str} > 0.5$, corresponding to the deformation of aggregates in Stage 2, along with the breaking of intra-aggregate contact (Figure \ref{fig:force_single_64_case1}(b)).
Finally, ${\langle Z \rangle}$ begins to increase with $\phi$ in Stage 3, where the hierarchical structure of the granular pile has been erased.

\section{Discussion}

\subsection{Impacts of dynamic pressure}

In this study, the top wall moves with a speed $v_{\rm wall}$.
Thus, the pressures calculated in our simulations ($P_{\rm top}$ and $P_{\rm bottom}$) are the sum of the static and dynamic terms.
The dynamic pressure, $P_{\rm dyn}$, can be approximately evaluated as follows:
\begin{equation}
P_{\rm dyn} \sim \frac{1}{2} \rho_{\rm bulk} {v_{\rm wall}}^{2} \approx 7 {\left( \frac{\phi}{0.15} \right)}^{-1}~{\rm Pa},
\label{eq:P_dyn}
\end{equation}
where $\rho_{\rm bulk} = \rho_{\bullet} \phi$ is the bulk density of hierarchical granular piles.
In our simulations, both $P_{\rm top}$ and $P_{\rm bottom}$ are higher than $10~{\rm Pa}$ for $\phi \ge 0.16$.
Thus, we expect that the pressures obtained from our numerical simulations reflect the static term.
In contrast, for $\phi \ll 0.16$ (gray area in Figures \ref{fig:P_phi_32} and \ref{fig:P_phi_64}), $P_{\rm top}$ and $P_{\rm bottom}$ in our simulations are comparable to or lower than $P_{\rm dyn}$, suggesting that the contribution of the dynamic term is not negligible in the low $\phi$ region.

It can be seen that $P_{\rm dyn}$ is proportional to ${v_{\rm wall}}^{2}$ and hence ${\tau_{\rm comp}}^{-2}$ (Equation (\ref{eq:P_dyn})).
Therefore, a smaller $\tau_{\rm comp}$ value must be chosen to investigate the compression curve of hierarchical granular piles with $\phi < 0.16$.
Additionally, the pressure in Stage 1 decreases with increasing $r_{\rm agg}$ (Equations (\ref{eq:P_S1}) and (\ref{eq:P_mid})), and simulations with a smaller $\tau_{\rm comp}$ value are also necessary to investigate the compression behavior of hierarchical granular piles consisting of aggregates with large radii, $r_{\rm agg} \gg 64 r_{\bullet}$.

\subsection{Dependence on $r_{\rm agg}$}
\label{sec:r_agg}

In Sections \ref{sec:results} and \ref{sec:single}, we present the results for $r_{\rm agg} = 32 r_{\bullet}$ and $64 r_{\bullet}$.
Our semi-analytic formula successfully reproduces the numerical results for both $r_{\rm agg} = 32 r_{\bullet}$ and $64 r_{\bullet}$; however, it should be noted that there is a limitation on the range of $r_{\rm agg}$ where the semi-analytic formula is applicable.

Figure \ref{fig:force_single_16} shows the force--displacement relationship for single aggregates with $r_{\rm agg} = 16 r_{\bullet}$.
We find that the force--displacement relationship cannot be expressed by Equation (\ref{eq:F_agg_wall}) when $r_{\rm agg} = 16 r_{\bullet}$ due to the prominent spike-like feature (especially for Run \#5).
For aggregates with small $r_{\rm agg}$ values, the number of particle--wall contacts is not large enough to apply a contact model for macroscopic plastic spheres \citep{doi:10.1080/14786443008565033}.
The large variation among simulation runs, characterized by the spike-like feature, corresponds to the small-size effect.
The interval of the spikes observed in Run \#5 is approximately $\sqrt{3} r_{\bullet}$, which corresponds to the inter-layer distance in a close-packed, face-centered cubic structure.
We note that the orientation of the layers is parallel to the walls during compression in Run \#5 (see Figure S4 in Supporting Information), and this peculiar condition might be the cause of the spike-like feature in the force--displacement relationship.

Figure \ref{fig:P_phi_16} shows the compression curve of a hierarchical granular pile with $r_{\rm agg} = 16 r_{\bullet}$.
Although our semi-analytic formula approximately matches the compression curve obtained from numerical simulations over a wide range of $\phi$, the formula cannot capture the transition from Stage 1 to Stage 2.
When $r_{\rm agg} \ge 32 r_{\bullet}$ (see Figure \ref{fig:semi-analytic}), the formulae for Stage 1 (blue curve) and Stage 2 (green curve) intersect at two points, and the compressive behavior changes at the intersection point at $\phi_{\rm str} \approx 0.45$ (red curve).
In contrast, for $r_{\rm agg} = 16 r_{\bullet}$, there are no intersection points between the formulae for Stage 1 and Stage 2.
Furthermore, the deviation of the fit of Stage 2 for $r_{\rm agg} = 16 r_{\bullet}$ is larger than for $r_{\rm agg} = 32 r_{\bullet}$ and $64 r_{\bullet}$, reflecting the spike-like feature and large deviation in the force--displacement relationship for single aggregates (see Figure \ref{fig:force_single_16}).

\subsection{Dependence on $N_{\rm agg}$}

In our simulations, the total number of particles is $N_{\rm tot} = N_{\rm agg} N_{\rm par}$.
The number of particles within each aggregate is given by $N_{\rm par} = \phi_{\rm agg} {( r_{\rm agg} / r_{\bullet} )}^{3}$, and $N_{\rm par} = 131072$ when $r_{\rm agg} = 64 r_{\bullet}$.
As discussed in Sections \ref{sec:semi-analytic} and \ref{sec:r_agg}, it is essential to choose sufficiently large $N_{\rm par}$ values for investigating the compressive behavior of hierarchical granular piles, especially in Stage 1 and Stage 2. 
The maximum $N_{\rm tot}$ value is limited by computational cost, so we set a moderately small value of $N_{\rm agg} = 200$ in this study.

The number of aggregates used in our simulation, $N_{\rm agg} = 200$, is not large enough to obtain precise results for the statistics of aggregate--aggregate contacts.
For instance, we investigated the distribution of the directions of aggregate--aggregate contacts in Figure \ref{fig:angle_phi_32}.
To discuss the deviation of the numerical results from the homogeneous distribution at $\phi_{\rm str} > 0.35$, simulations with a much larger $N_{\rm agg}$ will be needed in future studies.
Additionally, we will need to perform further simulations with a large $N_{\rm agg}$ value to obtain the compression curve for $\phi < 0.16$ (shaded region in Figures \ref{fig:P_phi_32} and \ref{fig:P_phi_64}).

\subsection{Dependence on $\phi_{\rm agg}$ and material parameters}

We fixed $\phi_{\rm agg} = 0.5$ and considered aggregates made of spherical ice particles with $r_{\bullet} = 0.1~{\mu}{\rm m}$.
It is evident that the force--displacement relationship (Equation (\ref{eq:F_agg_wall})) is significantly affected by the choice of $\phi_{\rm agg}$ and material parameters, including Young's modulus and surface energy.
We derived the material parameter dependence in Sections \ref{sec:single} and \ref{sec:stage1}.
The experimental data on the compression curves of both hierarchical and non-hierarchical granular piles composed of (sub)micron-sized silica grains have been compiled in Figure 4 of \citet{2024ApJ...974...76M}.
The compressive strengths of silica granular piles are comparable to our numerical results for ice granular piles for $\phi \le 0.3$, whereas the experimental data exceed our numerical results for $\phi > 0.3$.
This discrepancy may reflect differences in material composition and grain size, and we plan to investigate the compressive behavior of silica granular piles in future studies.

In contrast, the dependence on $\phi_{\rm agg}$ is not well understood, and further simulations are required.
For $\phi_{\rm agg} = 0.5$, the average coordination number within an aggregate is ${\langle Z_{\rm agg} \rangle} \approx 8$ at the initial condition.
For aggregates composed of frictionless spheres, the condition for jamming is ${\langle Z_{\rm agg} \rangle} \approx 6$ \citep{2003PhRvE..68a1306O}.
We speculate that the force--displacement relationship changes qualitatively depending on whether ${\langle Z_{\rm agg} \rangle} > 6$, and this hypothesis should be tested through numerical simulations.

It is important to note that, in reality, constituent particles exhibit tangential friction in natural aggregates.
The yield stress of porous aggregates composed of frictional particles has been extensively investigated in laboratory experiments \citep{2009ApJ...701..130G, 2017P&SS..149...14O, 2022A&A...664A.147O}.
As the next step, we will address numerical simulations of compression of hierarchical granular piles consisting of frictional particles.

\section{Conclusions}

\begin{figure}
\centering
\includegraphics[width = 0.8\columnwidth]{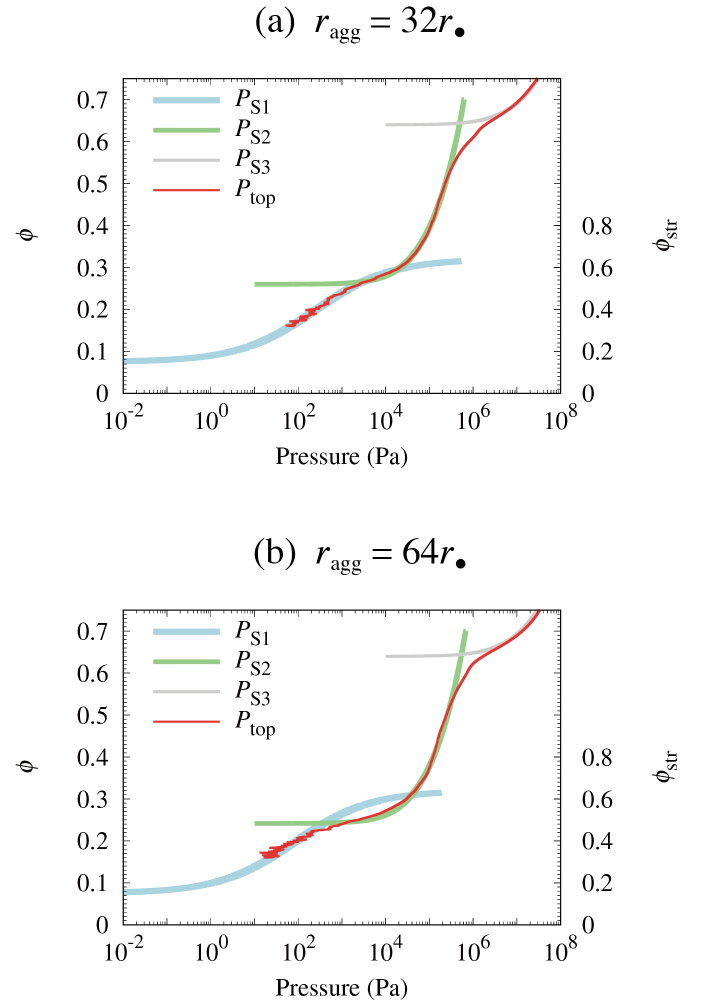}
\caption{
Semi-analytic fit of the compression curve of hierarchical granular piles.
The blue, green, and gray curves correspond to the semi-analytic formulae derived in this study (Equations (\ref{eq:P_S1}), (\ref{eq:P_S2}), and (\ref{eq:P_S3}), respectively).
The red curves represent the numerical results of $P_{\rm top}$ (see Figures \ref{fig:P_phi_32} and \ref{fig:P_phi_64}).
(a) $r_{\rm agg} = 32 r_{\bullet}$.
(b) $r_{\rm agg} = 64 r_{\bullet}$.
}
\label{fig:semi-analytic}
\end{figure}

\begin{figure}
\centering
\includegraphics[width = 0.8\columnwidth]{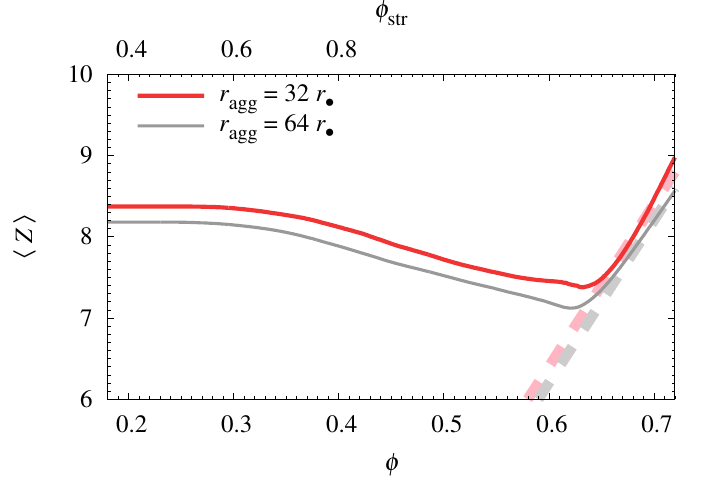}
\caption{
Average coordination number for interparticle contacts, ${\langle Z \rangle}$, as a function of $\phi$.
Dashed lines represent empirical fits: ${\langle Z \rangle} = 20 {( \phi - \phi_{3} )}$ and $\phi_{3} = 0.3 - 0.02 {( r_{\rm agg} / {( 32 r_{\bullet} )} )}^{-1}$.
}
\label{fig:Z_phi_S3}
\end{figure}

\begin{figure}
\centering
\includegraphics[width = 0.8\columnwidth]{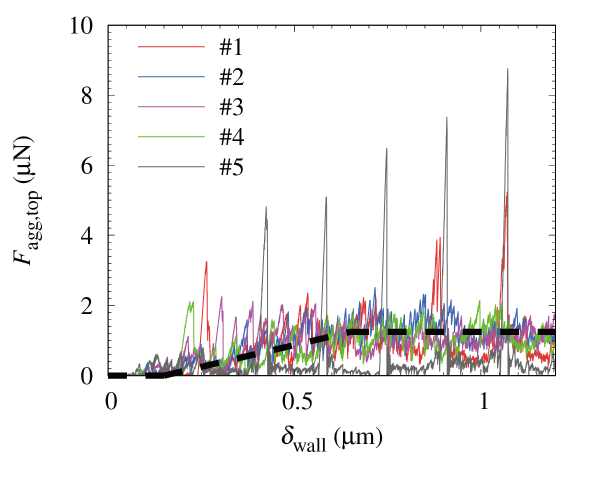}
\caption{
Same as Figure \ref{fig:force_single}, but for $r_{\rm agg} = 16 r_{\bullet}$.
}
\label{fig:force_single_16}
\end{figure}

\begin{figure}
\centering
\includegraphics[width = 0.8\columnwidth]{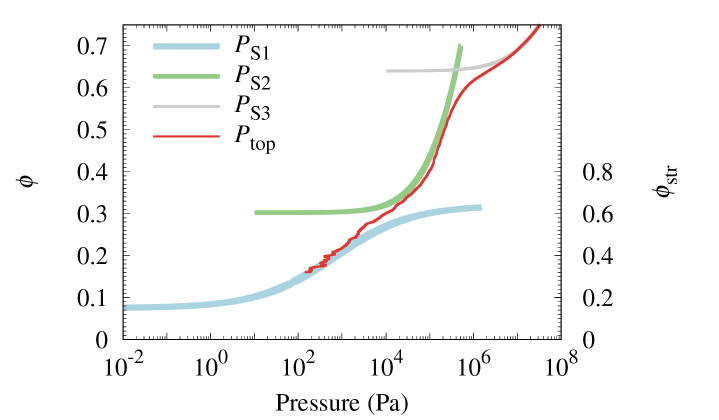}
\caption{
Same as Figure \ref{fig:semi-analytic}, but for $r_{\rm agg} = 16 r_{\bullet}$.
}
\label{fig:P_phi_16}
\end{figure}

In this study, we performed numerical compression tests of hierarchical granular piles consisting of tens of millions of particles.
Laboratory experiments have proposed that the packing structure of hierarchical granular piles evolves through a multi-step process \citep{2013Icar..226..111M, 2015Icar..257...33S, 2022Univ....8..381B}; however, this hypothesis has not yet been tested by numerical simulations.
We successfully reproduced and confirmed the multi-step evolution of the packing structure for the first time (Figure \ref{fig:semi-analytic}).
We also developed a quantitative semi-analytical model for the compression curve (see Section \ref{sec:semi-analytic}).
Our key findings are summarized as follows.

\begin{enumerate}
\item{The compression curve of hierarchical granular piles can be divided into three stages: (i) rearrangement of the aggregate packing structure, (ii) plastic deformation of small aggregates, and (iii) elastic deformation of constituent particles (see Figures \ref{fig:schematic_curve} and \ref{fig:semi-analytic}).}
\item{In Stage 1, structural changes of hierarchical granular piles occur without deformation of the aggregates.
The compressive curve for Stage 1 is approximately represented by a modified polytropic equation (Equation \ref{eq:P_S1}), where the friction for aggregate--aggregate rotation plays a key role.}
\item{In Stage 2, the average coordination number for aggregate--aggregate contacts exceeds 4 (see Figure \ref{fig:Z_phi_32} and Section \ref{sec:stage2}), and small aggregates can no longer rotate freely (Figure \ref{fig:v_phi}).
The plastic deformation of small aggregates within hierarchical granular piles causes repulsive forces to act between aggregates. 
The compressive curve for Stage 2 is approximately given by Equation (\ref{eq:P_S2}).}
\item{In Stage 3, small aggregates within hierarchical granular piles do not retain their original shapes.
The constituent particles in our simulations are elastic nanospheres, and repulsive forces act between the constituent particles when the volume filling factor of the hierarchical granular pile exceeds the jamming point.
The compressive curve for Stage 3 is approximately given by Equation (\ref{eq:P_S3}).}
\end{enumerate}

We considered granular piles consisting of ice particle with a radius of $0.1~{\mu}{\rm m}$; thus, our results could be directly applicable to the structural evolution of comets and other small icy bodies.
We will investigate the radial distribution of the internal density of comets using the model presented here.
Additionally, we can calculate the inter-aggregate contact area as a function of the applied pressure.
Since the thermal conductivity within hierarchical granular piles is controlled by the inter-aggregate contact area \citep{2023MNRAS.521.4927A}, our semi-analytic formula is essential for modelling heat conduction within comets.
We will revisit the early thermal evolution of small icy bodies \citep{2022MNRAS.514.3366M, 2024PASJ...76..130A}.
We also plan to investigate the role of hierarchical granular piles in rocks and soils within fault zone, using large-scale DEM simulations for earthquakes \citep{2024Tectp.87430230F} and landslides \citep{2024CGeot.16505855C}.

\clearpage

\section*{Declarations}

\subsection*{Availability of data and material}

This manuscript is based on results computed from a numerical model, not on data obtained elsewhere by ourselves or others.
These numerical results are freely accessible at {doi: 10.5281/zenodo.13999795}.

\subsection*{Competing interests}

Not applicable.

\subsection*{Funding}

JSPS KAKENHI Grant (JP24K17118 and JP24H00279).

\subsection*{Authors' contributions}

SA developed the idea for the study.
MF and SA performed numerical simulations with support from DN.
All of the authors contributed to the interpretation of the results.

\section*{Acknowledgments}

We thank two anonymous reviewers for constructive comments.
This study was supported by the Earth Simulator project of JAMSTEC (Development of advanced particle simulation code).
This study was supported by JSPS KAKENHI Grant (JP24K17118 and JP24H00279).

\clearpage


\clearpage

\clearpage

\clearpage

\clearpage

\clearpage

\clearpage

\clearpage

\clearpage

\clearpage

\clearpage

\clearpage

\clearpage

\clearpage

\clearpage

\clearpage

\clearpage

\clearpage

\clearpage

\clearpage

\clearpage

\clearpage

\newpage

\begin{figure}
\centering
\includegraphics[width=\textwidth, page=1]{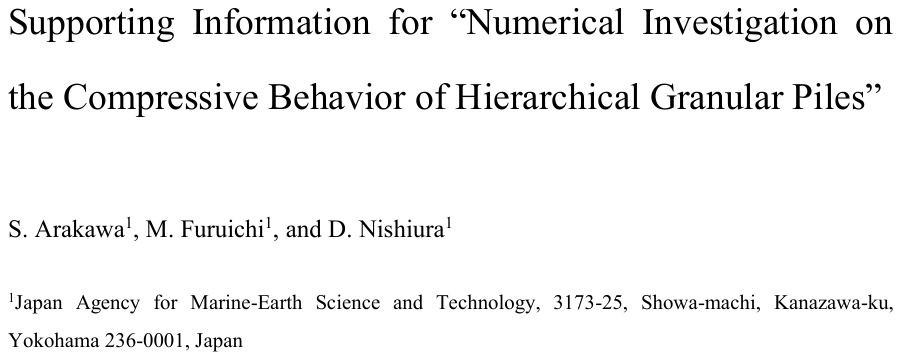}
\caption*{}
\end{figure}

\begin{figure}
\centering
\includegraphics[width=\textwidth, page=2]{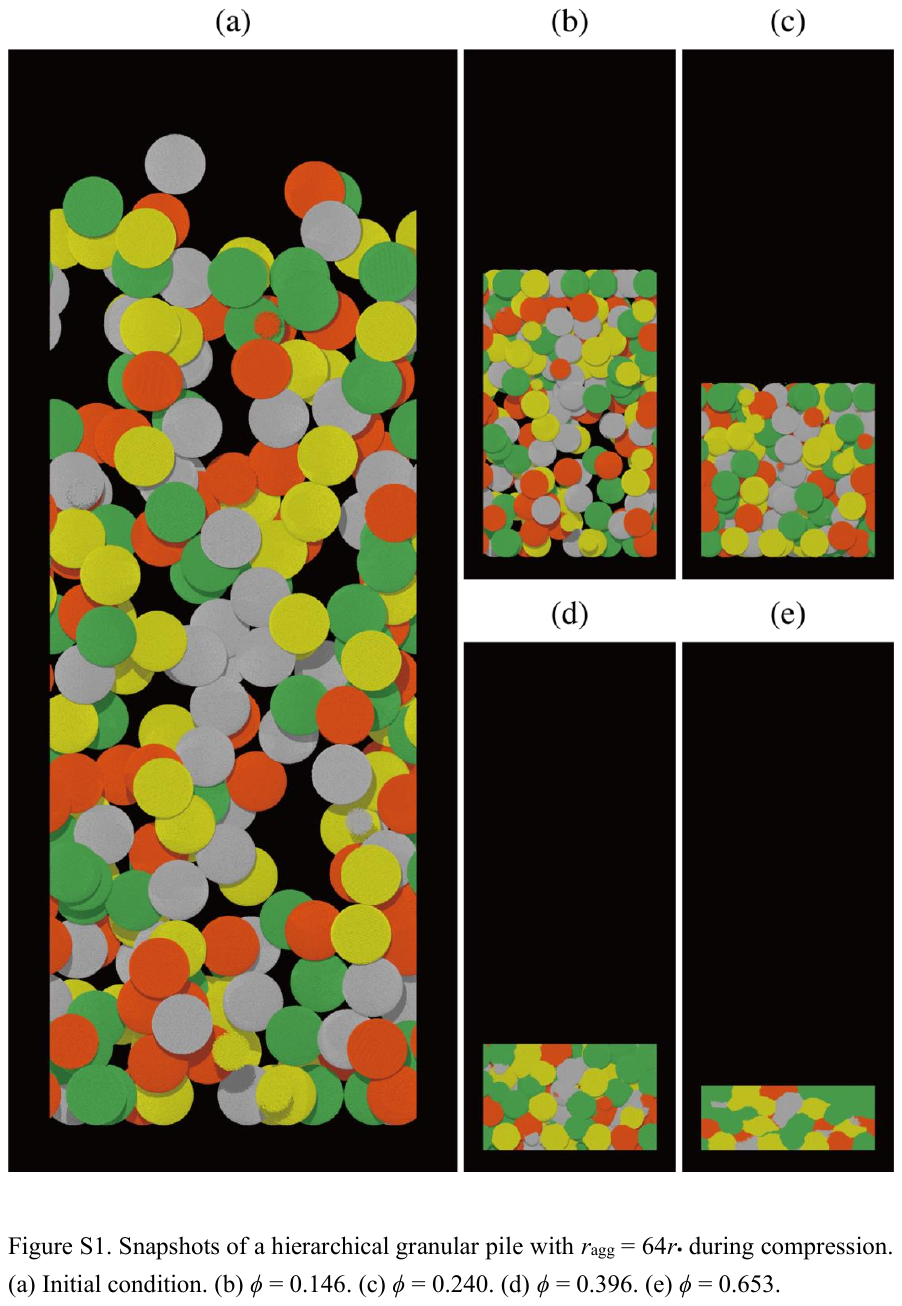}
\caption*{}
\end{figure}

\begin{figure}
\centering
\includegraphics[width=\textwidth, page=3]{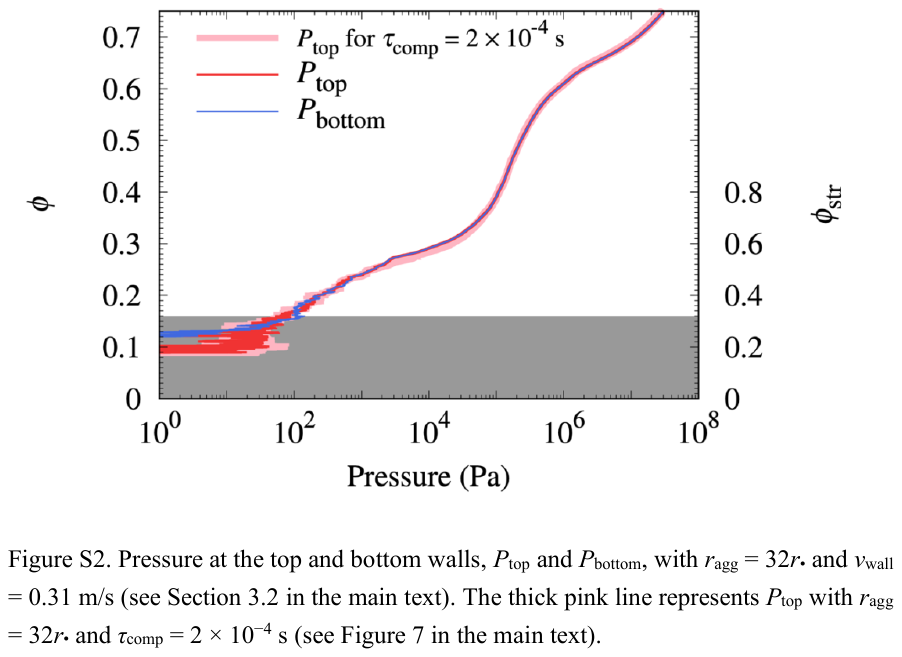}
\caption*{}
\end{figure}

\begin{figure}
\centering
\includegraphics[width=\textwidth, page=4]{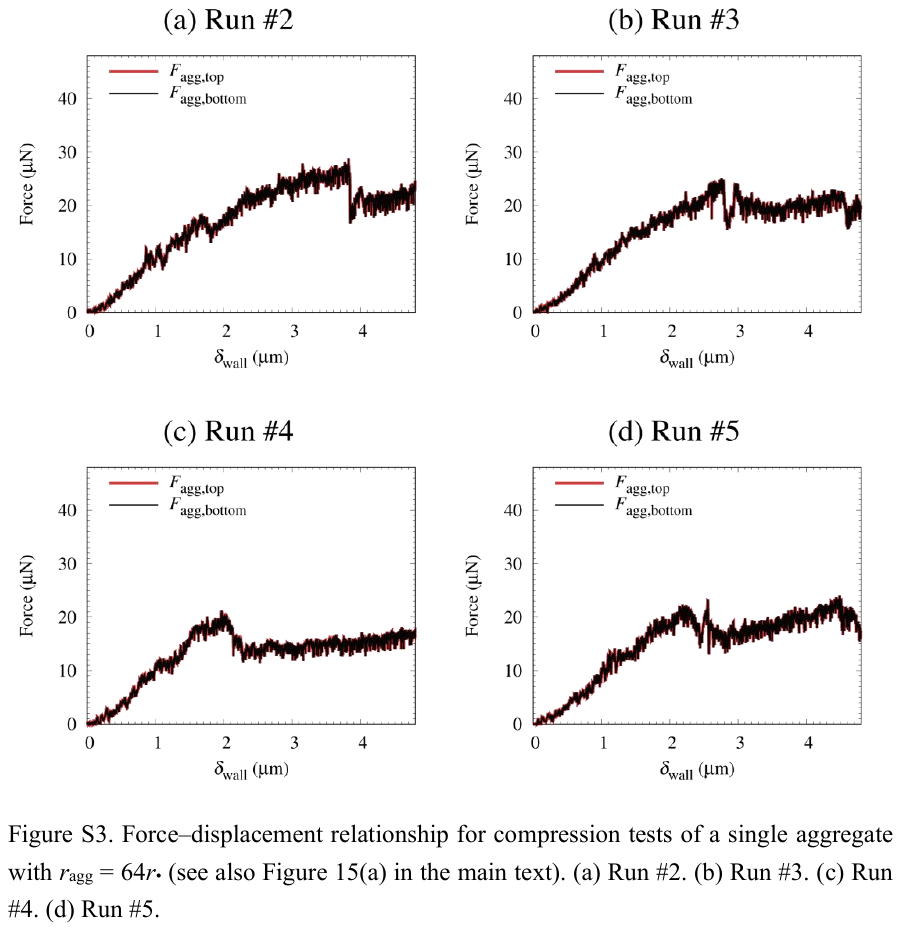}
\caption*{}
\end{figure}

\begin{figure}
\centering
\includegraphics[width=\textwidth, page=5]{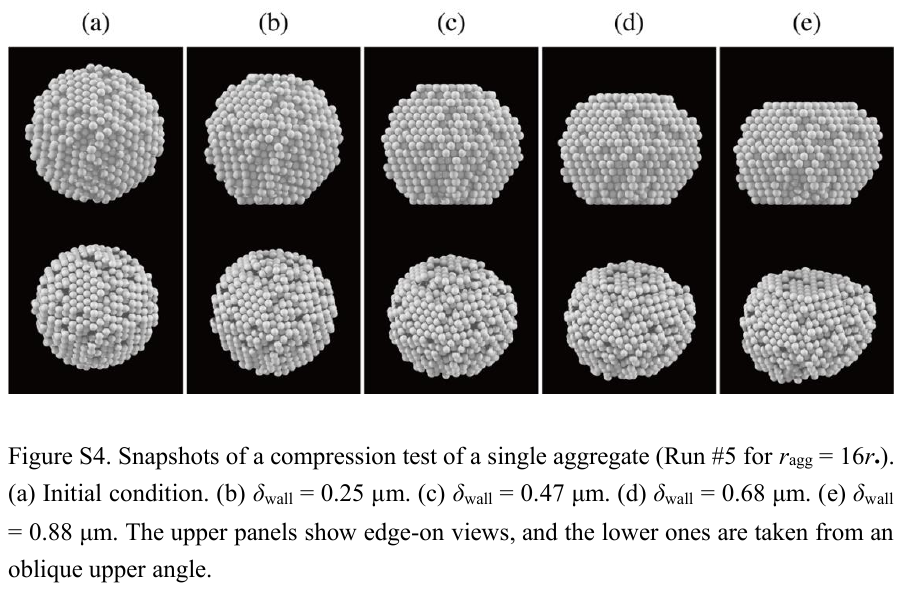}
\caption*{}
\end{figure}

\end{document}